\documentclass[a4paper,11pt]{article}
\pdfoutput=1 

\usepackage{jheppub}
\usepackage{amsfonts}
\usepackage{amssymb}
\usepackage{comment}
\usepackage{epsfig}
\usepackage{graphicx}
\usepackage{amsmath}
\usepackage{tabu}
\usepackage{subfig}
\usepackage{float}
\usepackage{epstopdf}
\usepackage{multirow}
\usepackage{lscape}
\usepackage[utf8]{inputenc}

\newcommand{\be}{\begin{equation}}
\newcommand{\ee}{\end{equation}}
\def\bea{\begin{eqnarray}}
\def\eea{\end{eqnarray}}
\def\KK{{\scriptscriptstyle KK}}
\def\DM{{\scriptscriptstyle DM}}
\def\E3{{\scriptscriptstyle E3}}
\def\D7{{\scriptscriptstyle D7}}

\def\ALP{{\scriptscriptstyle ALP}}
\def\QCD{{\scriptscriptstyle QCD}}
\def\vo{\mathcal{V}}
\def\nn{\nonumber}
\def\mc{\mathcal}

\title{The 3.5 keV Line from Stringy Axions}

\author[1,2,3]{Michele Cicoli,}
\author[1,2]{Victor A. Diaz,}
\author[1]{Veronica Guidetti,}
\author[4,5]{Markus Rummel}

\affiliation[1]{Dipartimento di Fisica e Astronomia, Universit\`{a} di Bologna,\\via Irnerio 46, 40126 Bologna, Italy}
\affiliation[2]{INFN, Sezione di Bologna, via Irnerio 46, 40126 Bologna, Italy}
\affiliation[3]{ICTP, Strada Costiera 11, Trieste 34151, Italy}
\affiliation[4]{Department of Physics \& Astronomy, McMaster University\\ \qquad 1280 Main Street West, Hamilton ON, Canada}
\affiliation[5]{Perimeter Institute for Theoretical Physics\\ \qquad 31 Caroline Street North, Waterloo ON, Canada}

\emailAdd{mcicoli@ictp.it}
\emailAdd{diaz@bo.infn.it}
\emailAdd{veronica.guidetti@studio.unibo.it}
\emailAdd{rummelm@mcmaster.ca}

\abstract{An interesting result in particle astrophysics is the recent detection of an unexplained $3.5$ keV line from galaxy clusters. A promising model, which can explain the morphology of the signal and its non-observation in dwarf spheroidal galaxies, involves a $7$ keV dark matter particle decaying into a pair of ultra-light axions that convert into photons in the magnetic field of the clusters. Given that light axions emerge naturally in 4D string vacua, in this paper we present a microscopic realisation of this model within the framework of type IIB flux compactifications. Dark matter is a local closed string axion which develops a tiny mass due to subdominant poly-instanton corrections to the superpotential and couples via kinetic mixing to an almost massless open string axion living on a D3-brane at a singularity. The interaction of this ultra-light axion with photons is induced by $U(1)$ kinetic mixing. After describing the Calabi-Yau geometry and the brane set-up, we discuss in depth moduli stabilisation, the resulting mass spectrum and the strength of all relevant couplings.}

\begin{document} 
\maketitle
\flushbottom

\section{Introduction}
\label{Intro}

Recently several studies have shown the appearance of a photon line at $E\sim3.5$ keV, based on stacked X-ray data from galaxy clusters and the Andromeda galaxy \cite{Bulbul,Boyarsky}. The line has been detected in galaxy clusters by the X-ray observatories XMM-Newton, Chandra and Suzaku~\cite{Urban:2014yda,Franse:2016dln} and in Andromeda with XMM-Newton~\cite{Boyarsky}. The Hitomi satellite would have been able to study the 3.5 keV line with unprecedented energy resolution. However, unfortunately Hitomi was lost after only a few weeks in operation and the limited exposure time on the Perseus cluster only allows to put upper bounds on the $3.5$ keV line which are consistent with the detection of the other satellites~\cite{Aharonian:2016gzq}. The findings of~\cite{Bulbul,Boyarsky} have inspired further searches in other astrophysical objects such as the galactic center~\cite{Riemer-Sorensen:2014yda,Jeltema:2014qfa,Boyarsky:2014ska,Carlson:2014lla}, galaxies~\cite{Anderson:2014tza}, dwarfs~\cite{Malyshev:2014xqa,Jeltema:2015mee,Ruchayskiy:2015onc} and other galaxy clusters~\cite{Iakubovskyi:2015dna,Hofmann:2016urz}.\footnote{For a summary of observations and models on the $3.5$ keV line see \cite{Iakubovskyi:2015wma}.}  Currently, a compelling standard astrophysical explanation, e.g. in terms of atomic lines of the (cluster) gas is lacking.\footnote{See however \cite{Jeltema:2014qfa,Gu:2015gqm}.} This gives rise to the possibility that the $3.5$ keV line is a signal related to dark matter (DM) physics.

A much explored model is that of dark matter decay, e.g. a sterile neutrino with mass $m_\DM \sim 7$ keV decaying into an active neutrino and a photon~\cite{Dodelson:1993je,Asaka:2005pn}. In this case, the photon flux from an astrophysical object is solely determined by the lifetime of the dark matter particle and the dark matter column density. The width of the line is due to Doppler broadening. There are several observational tensions, if one wants to explain the (non-)observation of the $3.5$ keV line in currently analysed astrophysical objects. Most prominently, these are:
\begin{itemize}
\item Non-observation of the $3.5$ keV line from dwarf spheroidal galaxies~\cite{Malyshev:2014xqa,Jeltema:2015mee,Ruchayskiy:2015onc}. The dark matter density of these objects is rather well known and the X-ray background is low, making dwarf spheroidals a prime target for detecting decaying dark matter.
\item Non-observation of the $3.5$ keV line from spiral galaxies~\cite{Anderson:2014tza}, where again the X-ray background is low. According to the dark matter estimates of~\cite{Anderson:2014tza}, the non-observation of a $3.5$ keV signal from spiral galaxies excludes a dark matter decay origin of the $3.5$ keV line very strongly at $11\, \sigma$.
\item The radial profile of the $3.5$ keV line in the Perseus cluster peaks on shorter scales than the dark matter profile, rather following the gas profile than the dark matter profile~\cite{Bulbul,Carlson:2014lla}. However, the observed profile with Suzaku is only in mild tension with the dark matter profile~\cite{Franse:2016dln}.
\end{itemize}
These tensions, even though they could be potentially explained by uncertainties in the dark matter distributions in these objects~\cite{Iakubovskyi:2015wma}, motivate different dark matter models than direct dark matter decay into a pair of $3.5$~keV photons.

A dark matter model that is consistent with all the present (non-)observations was given in \cite{Cicoli:2014bfa}. A dark matter particle with mass $m_\DM \sim 7$ keV decays into an almost massless ($m_\ALP \lesssim 10^{-12}$ eV) axion-like particle (ALP) with energy $3.5$ keV which successively converts into $3.5$ keV photons that are finally observed. Compared to direct dark matter decay into photons, the observed photon flux does not depend just on the dark matter column density, but also on the probability for ALPs to convert into photons. This is determined by the size and coherence scale of the magnetic field and the electron density in e.g. a galaxy cluster.

The $3.5$ keV emission is stronger in astrophysical regions with relatively large and coherent magnetic field. This is verified by the experimental fact that cool core clusters like the Perseus cluster have stronger magnetic fields than non-cool core clusters and also a higher $3.5$ keV flux is observed from such an object. Furthermore, the fact that central regions of a cool core cluster host particularly strong magnetic fields explains the radial morphology of the $3.5$ keV flux from Perseus as the signal comes disproportionally from the central region of the cluster. The model has made the prediction that galaxies can only generate a non-negligible $3.5$ keV photon flux if they are spiral and edge-on as for instance the Andromeda galaxy \cite{Cicoli:2014bfa}. In this case, the full length of the regions with regular magnetic field can be used efficiently for ALP to photon conversion. These predictions agree with the experimental results of non-observation of the $3.5$ keV signal from generic (edge-on and face-on) spiral galaxies and dwarf galaxies~\cite{14101867}.\footnote{Despite the successful interpretation of all these observations, this model would not be able to explain the dip around $3.5$ keV in the Perseus AGN spectrum which might arise from Chandra data \cite{Conlon:2016lxl}.}

Given that the 4D low-energy limit of string compactifications generically leads to several light ALPs \cite{Svrcek:2006yi, Conlon:2006tq, Arvanitaki:2009fg, Acharya:2010zx, Cicoli:2012sz}, it is natural to try to embed the model of \cite{Cicoli:2014bfa} in string theory. This is the main goal of this paper where we focus in particular on type IIB flux compactifications where moduli stabilisation has already been studied in depth. 

In 4D string models, ALPs can emerge either as closed string modes arising from the dimensional reduction of 10D anti-symmetric forms or as phases of open string modes charged under anomalous $U(1)$ symmetries on stacks of D-branes \cite{Svrcek:2006yi, Conlon:2006tq, Arvanitaki:2009fg, Acharya:2010zx, Cicoli:2012sz}. Some of these modes can be removed from the low-energy spectrum by the orientifold projection which breaks $N=2$ supersymmetry down to $N=1$, others can be eaten up by anomalous $U(1)$'s via the Green-Schwarz mechanism for anomaly cancellation or can become as heavy as the gravitino if the corresponding saxions are stabilised by the same non-perturbative effects which give mass to the axions. However the axions enjoy a shift symmetry which is broken only at non-perturbative level. Therefore when the corresponding saxions are frozen by perturbative corrections to the effective action, the axions remain exactly massless at this level of approximation. They then develop a mass via non-perturbative effects which are however exponentially suppressed with respect to perturbative corrections. Hence whenever perturbative contributions to the effective scalar potential play a crucial r\^ole for moduli stabilisation, the axions are exponentially lighter than the associated saxions \cite{Allahverdi:2014ppa}. Notice that this case is rather generic in string compactifications for two main reasons: $(i)$ if the background fluxes are not tuned, non-perturbative effects are naturally subleading with respect to perturbative ones; $(ii)$ it is technically difficult to generate non-perturbative contributions to the superpotential which depend on all moduli (because of possible extra fermionic zero modes \cite{Blumenhagen:2009qh}, chiral intersections with the visible sector \cite{Blumenhagen:2007sm} or non-vanishing gauge fluxes due to Freed-Witten anomaly cancellation \cite{Freed:1999vc}). 

\begin{figure}
\centerline{\psfig{file=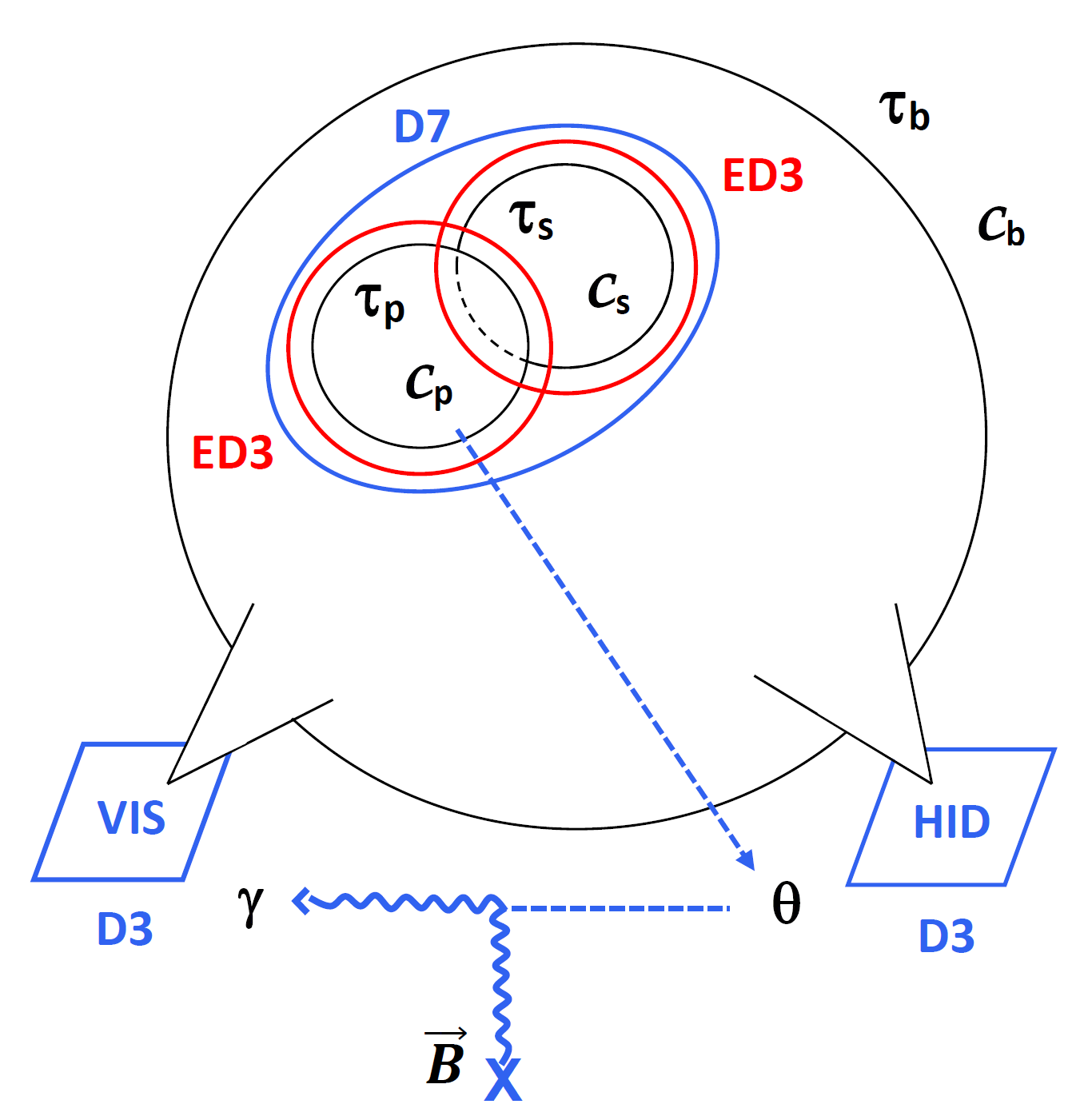,width=10.0cm}}
\caption{Pictorial view of our setup: a stack of D7-branes wraps the combination $\tau_s+\tau_p$, two ED3-instantons wrap respectively the rigid cycle $\tau_s$ and the Wilson divisor $\tau_p$ while two stacks of D3-branes at singularities support the visible and a hidden sector. The DM particle is the closed string axion $c_p$ which acquires a $7$ keV mass due to tiny poly-instanton effects and decays to the ultra-light open string ALP $\theta$ that gives the $3.5$ keV line by converting into photons in the magnetic field of galaxy clusters.}
\label{Fig}
\end{figure}

String compactifications where some moduli are fixed by perturbative effects are therefore perfect frameworks to derive models for the $3.5$ keV line with light ALPs which can behave as either the $7$ keV decaying DM particle or as the ultra-light ALP which converts into photons. The main moduli stabilisation mechanism which exploits perturbative corrections to the K\"ahler potential is the LARGE Volume Scenario (LVS) \cite{Balasubramanian:2005zx, Conlon:2005ki, Cicoli:2008va}. We shall therefore present an LVS model with the following main features (see Fig. \ref{Fig} for a pictorial view of our microscopic setup):
\begin{itemize}
\item The underlying Calabi-Yau (CY) manifold is characterised by $h^{1,1}=5$ K\"ahler moduli $T_i = \tau_i+{\rm i} c_i$ where the $c_i$'s are closed string axions while the $\tau_i$'s control the volume of $5$ different divisors: a large four-cycle $D_b$, a rigid del Pezzo four-cycle $D_s$ which intersects with a `Wilson divisor' $D_p$ ($h^{0,1}(D_p)=1$ and $h^{0,2}(D_p)=0$) and two non-intersecting blow-up modes $D_{q_1}$ and $D_{q_2}$.

\item The two blow-up modes $D_{q_1}$ and $D_{q_2}$ shrink down to zero size due to D-term stabilisation and support D3-branes at the resulting singularities. These constructions are rather promising to build a semi-realistic visible sector with SM-like gauge group, chiral spectrum and Yukawa couplings \cite{Aldazabal:2000sa, Conlon:2008wa}. If $D_{q_1}$ and $D_{q_2}$ are exchanged by the orientifold involution, the visible sector features two anomalous $U(1)$ symmetries (this is always the case for any del Pezzo singularity) \cite{Cicoli:2012vw, Cicoli:2013mpa, Cicoli:2013cha}, while if the two blow-up modes are separately invariant, one of them supports the visible sector and the other a hidden sector \cite{Wijnholt:2007vn, Cicoli:2017shd}. Each of the two sectors is characterised by a single anomalous $U(1)$ factor. 

\item A smooth combination of $D_s$ and $D_p$ is wrapped by a stack of D7-branes which give rise to string loop corrections to the K\"ahler potential $K$ \cite{Berg:2005ja, Berg:2007wt, Cicoli:2007xp}. Moreover, non-vanishing world-volume fluxes generate moduli-dependent Fayet-Iliopoulos (FI) terms \cite{Dine:1987xk, Dine:1987gj}. An ED3-instanton wraps the rigid divisor $D_s$ and generates standard $T_s$-dependent non-perturbative corrections to the superpotential $W$. A second ED3-instanton wraps the Wilson divisor $D_p$. Due to the presence of Wilson line modulini, this ED3-instanton contributes to the superpotential only via $T_p$-dependent poly-instanton effects \cite{Blumenhagen:2008ji, Blumenhagen:2012kz}.

\item At leading order in an inverse volume expansion, the moduli are fixed supersymmetrically by requiring vanishing D- and F-terms. These conditions fix the dilaton and the complex structure moduli in terms of three-form flux quanta together with the blow-up modes $\tau_{q_1}$ and $\tau_{q_2}$ in terms of charged open string fields, and hidden matter fields on the D7-stack in terms of $\tau_p$.

\item Quantum corrections beyond tree-level break supersymmetry and stabilise most of the remaining flat directions: $\alpha'$ corrections to $K$ \cite{Becker:2002nn} and single non-perturbative corrections to $W$ \cite{Kachru:2003aw} fix $\tau_b$, $\tau_s$ and $c_s$, while soft supersymmetry breaking mass terms and $g_s$ loop corrections to $K$ fix $\tau_p$.

\item Subdominant $T_p$-dependent poly-instanton corrections to $W$ stabilise the local closed string axion $c_p$ while a highly suppressed $T_b$-dependent non-perturbative superpotential fixes the bulk closed string axion $c_b$. Sequestered soft term contributions stabilise instead the radial component of $U(1)$-charged matter fields $C=|C|\,e^{{\rm i}\,\theta}$ living on the D3-brane stacks.

\item Both $c_b$ and $c_p$ are exponentially lighter than the gravitino, and so could play the r\^ole of the decaying DM particle with $m_\DM\sim 7$ keV.
On the other hand the ultra-light ALP with $m_\ALP\lesssim 10^{-12}$ eV which converts into photons is given by the open string phase $\theta$. Notice that if $D_{q_1}$ and $D_{q_2}$ are identified by the orientifold involution, there are two open string phases in the visible sector: one behaves as the standard QCD axion, which is however heavier than $10^{-12}$ eV, and the other is the ultra-light ALP $\theta$. If instead $D_{q_1}$ and $D_{q_2}$ are separately invariant under the involution, $\theta$ is an open string axion belonging to a hidden sector. 

\item The coupling of the closed string axions $c_b$ and $c_p$ to the open string ALP $\theta$ is induced by kinetic mixing due to non-perturbative corrections to the K\"ahler potential. However we shall show that the scale of the induced DM-ALP coupling can be compatible with observations only if the DM candidate is the local closed string axion $c_p$.

\item If the ultra-light ALP $\theta$ belongs to the hidden sector, its coupling to ordinary photons can be induced by $U(1)$ kinetic mixing which gets naturally generated by one-loop effects \cite{Abel:2008ai}. Interestingly, the strength of the resulting interaction can easily satisfy the observational constraints if the open string sector on the D3-brane stack is both unsequestered and fully sequestered from the sources of supersymmetry breaking in the bulk.

\item The branching ratio for the direct axion DM decay into ordinary photons is negligible by construction since it is induced by kinetic mixing between Abelian gauge boson on the D7-stack and ordinary photons on the D3-stack which gives rise to an interaction controlled by a scale which is naturally trans-Planckian.
\end{itemize}

This paper is organised as follows. In Sec. \ref{Pheno} we first discuss the phenomenology of the dark matter to ALP to photon model for the $3.5$ keV line and its observational constraints, and then we describe how these phenomenological conditions turn into precise requirements on the Calabi-Yau geometry, the brane setup and gauge fluxes, the 4D fields which can successfully play the r\^ole of either the DM particle or the ultra-light ALP, the form of the various interactions and the resulting low-energy 4D supergravity. Sec. \ref{ModStab} provides a thorough discuss of moduli stabilisation showing how different sources of corrections to the effective action can fix all closed string moduli and the $U(1)$-charged open string modes. In Sec. \ref{MassCoupl} we first derive the expressions for the canonically normalised fields and their masses and then we use these results to work out the strength of the DM-ALP coupling before presenting our conclusions in Sec. \ref{Concl}. Several technical details are relegated to App. \ref{App}.

\section{Phenomenology and microscopic realisation}
\label{Pheno}

In this section we first discuss the observational constraints of the model of \cite{Cicoli:2014bfa} for the $3.5$ keV line, and we outline the main phenomenological features of our embedding in LVS type IIB flux compactifications. We then provide the technical details of the microscopic realisation of the DM to ALP to photon model for the $3.5$ keV line. We start by illustrating the geometry of the underlying Calabi-Yau compactification manifold. We then present the brane setup and gauge fluxes, and we finally describe the main features of the resulting low-energy 4D effective field theory.

\subsection{Observational constraints} 
\label{PhenoConstr}

The effective Lagrangian of the dark matter to ALP to photon model for the $3.5$ keV line can be described as follows:
\bea
\mc{L}&= &-\frac14 F^{\mu\nu}F_{\mu\nu} -\frac{a_\ALP}{4M}\,F^{\mu\nu}\tilde{F}_{\mu\nu} +\frac12 \partial_\mu a_\ALP \partial^\mu a_\ALP
-\frac12 m_\ALP^2 a_\ALP^2 \nn \\
&+& \frac{a_\DM}{\Lambda} \,\partial_\mu a_\ALP \partial^\mu a_\ALP + \frac12 \partial_\mu a_\DM\partial^\mu a_\DM -\frac12 m_\DM^2 a_\DM^2\,,  
\label{effL} 
\eea
where $a_\ALP$ is an ALP with mass $m_\ALP$ that converts into photons in astrophysical magnetic fields via the coupling suppressed by $M$. $a_\DM$ is a pseudoscalar which is the dark matter particle with mass $m_\DM \sim 7$ keV. It decays via the kinetic mixing term in (\ref{effL}) with characteristic scale $\Lambda$. In order for ALP-photon conversion to be efficient in galaxy cluster magnetic field environments, we require $m_\ALP \lesssim 10^{-12}$ eV which is the characteristic energy scale of the electron-photon plasma \cite{Cicoli:2014bfa}. Otherwise, the ALP to photon conversion is suppressed by $\sim (10^{-12} \,{\rm eV} / m_\ALP)^4$. Therefore $a_\ALP$ is too light to be the standard QCD axion but it has instead to be a stringy axion-like particle.

The observed photon flux at an X-ray detector is given by:
\be
F_{DM\rightarrow \,a_\ALP \rightarrow \,\gamma} \propto \Gamma_{a_\DM\rightarrow a_\ALP a_\ALP}\,P_{a_\ALP \rightarrow \gamma}\, \rho_\DM\,,
\label{eq:flux}
\ee
where $\rho_\DM$ is the dark matter column density and:
\be
\Gamma_{a_\DM\rightarrow a_\ALP a_\ALP}=\frac{1}{32\pi}\frac{m_\DM^3}{\Lambda^2}\,,
\ee
is the dark matter decay rate and $P_{a_\ALP \rightarrow \gamma}$ is the ALP to photon conversion probability. It is given as $P_{a_\ALP \rightarrow \gamma} \propto M^{-2}$ and furthermore depends on the electron density in the plasma, the energy of the ALP/photon, the coherence length and the strength of the magnetic field. Hence, $F_{DM\rightarrow a_\ALP \rightarrow \gamma} \propto \Lambda^{-2} M^{-2}$. For the ALP to photon conversion conditions in the Perseus cluster magnetic field, the observed $3.5$~keV flux then implies \cite{Cicoli:2014bfa}:
\be
\label{LambdatimesM}
\Lambda\cdot M\sim 7\cdot 10^{28} \,\,{\rm GeV}^2\,.
\ee
The scales $M$ and $\Lambda$ are subject to certain constraints. There is a lower bound $M \gtrsim 10^{11}$ GeV from observations of SN1987A~\cite{Brockway:1996yr,Grifols:1996id,Payez:2014xsa}, the thermal spectrum of galaxy clusters \cite{Conlon:2015uwa} and active galactic nuclei~\cite{Berg:2016ese,Marsh:2017yvc,Conlon:2017qcw}. This lower bound implies an upper bound on $\Lambda$ via~\eqref{LambdatimesM}. To get sufficiently stable dark matter, we assume that the dark matter particle has a lifetime larger than the age of the universe, i.e. $\Lambda \gtrsim 5 \cdot 10^{12}$ GeV. This implies an upper bound on $M$ via (\ref{LambdatimesM}). To summarise, the parameters $M$ and $\Lambda$ have to satisfy (\ref{LambdatimesM}) together with the following phenomenological constraints:
\be
10^{11} \,\,{\rm GeV} \lesssim\, M \lesssim 10^{16} \,\,{\rm GeV}\,, \qquad
5\cdot 10^{12}\,\,{\rm GeV} \lesssim \,\Lambda \lesssim 7\cdot 10^{17}\,\,{\rm GeV}\,.
\label{LambdaMwindow}
\ee
Notice that ultra-light ALPs with intermediate scale couplings to photons will be within the detection reach of helioscope experiments like IAXO \cite{Irastorza:2013dav} and potentially light-shining-through-a-wall experiments like ALPS~\cite{Spector:2016vwo}.

\subsection{Phenomenological features}
\label{PhenoFeatures}

The phenomenological requirements for a viable explanation of the $3.5$ keV line from dark matter decay to ALPs which then convert into photons, can be translated into precise conditions on the topology and the brane setup of the microscopic realisation. We shall focus on type IIB flux compactifications where moduli stabilisation has already been studied in depth. According to (\ref{LambdatimesM}) and (\ref{LambdaMwindow}), we shall focus on the parameter space region where the DM to ALP coupling is around the GUT/Planck scale, $\Lambda\sim 10^{16}$-$10^{18}$ GeV, whereas the ALP to photon coupling is intermediate: $M\sim 10^{11}$-$10^{13}$ GeV. This region is particularly interesting since an ALP with this decay constant could also explain the diffuse soft X-ray excess from galaxy cluster via axion-photon conversion in the cluster magnetic field \cite{Conlon:2013txa}. This phenomenological requirement, together with the observation that $m_\DM\sim 10$ keV while $m_\ALP \lesssim 10^{-12}$ eV, sets the following model building constraints:
\begin{itemize}
\item \textbf{ALP as an open string axion at a singularity:} From the microscopic point of view, $a_\ALP$ can be either a closed or an open string axion. In the case of closed string axions, $a_\ALP$ could be given by the reduction of $C_4$ on orientifold-even four-cycles or by the reduction of $C_2$ on two-cycles duals to orientifold-odd four-cycles. As explained in \cite{Conlon:2006tq, Cicoli:2012sz} and reviewed in App. \ref{AxEFT}, since axions are the imaginary parts of moduli, $T_i = \tau_i + {\rm i}\,c_i$ ($c_i$ is a canonically unnormalised axion), whose interaction with matter is gravitational, they tend to be coupled to photons with Planckian strength. However this is true only for bulk axions which have $M\simeq M_p$, while the coupling to photons of local axions, associated to blow-up modes of point-like singularities, is controlled by the string scale: $M\simeq M_s$. $M_s\sim M_p/\sqrt{\vo}$ can be significantly lower than $M_p$ if the volume of the extra dimensions in string units $\vo$ is very large, and so local closed string axions could realise $M\sim M_s \sim 10^{11}$-$10^{13}$ GeV.

A moduli stabilisation scheme which leads to an exponentially large $\vo$ is the LARGE Volume Scenario \cite{Balasubramanian:2005zx, Conlon:2005ki, Cicoli:2008va} whose simplest realisation requires a Calabi-Yau volume of the form:
\be
\vo=\tau_b^{3/2}-\tau_s^{3/2}\,.
\label{simplestLVS}
\ee
The moduli are fixed by the interplay of the leading order $\alpha'$ correction to the K\"ahler potential and non-perturbative effects supported on the rigid cycle $\tau_s$. The decay constant of the axionic partner of $\tau_s$, which we denote as $c_s$, is set by the string scale, $M\sim M_s$, but this mode develops a mass of order the gravitino mass $m_{c_s}\sim m_{3/2}\sim M_p/\vo$. The large divisor $\tau_b$ is lighter than the gravitino due to the underlying no-scale structure of the 4D effective field theory, $m_{\tau_b}\sim m_{3/2}/\sqrt{\vo}$, but it has to be heavier than about $50$ TeV in order to avoid any cosmological moduli problem. Hence the local axion $c_s$ is much heavier than $10^{-12}$ eV, and so cannot play the r\^ole of $a_\ALP$. Moreover, the bulk axion $c_b$ cannot be the desired ALP as well since, even if it is almost massless, its coupling scale to photons would be too high: $M\sim M_p$. 

We are therefore forced to consider an open string axion realisation for $a_\ALP$. Anomalous $U(1)$ factors appear ubiquitously in both D7-branes wrapped around four-cycles in the geometric regime and in D3-branes at singularities. In the process of anomaly cancellation, the $U(1)$ gauge boson becomes massive by eating up an axion \cite{Green:1984sg}. As explained in \cite{Allahverdi:2014ppa}, the combination of axions which gets eaten up is mostly given by an open string axion for D7-branes and by a closed string axion for D3-branes. The resulting low-energy theory below the gauge boson mass, features a global $U(1)$ which is an ideal candidate for a Peccei-Quinn like symmetry. In the case of D3-branes at singularities, the resulting D-term potential looks schematically like:
\be
V_D = g^2 \left( q |\hat{C}|^2-\xi \right)^2\,,
\ee
where we focused just on one canonically normalised charged matter field $\hat{C} = |\hat{C}|\,e^{{\rm i}\, \theta}$ whose phase $\theta$ can play the r\^ole of an axion with decay constant set by the VEV of the radial part $|\hat{C}|$. The FI term $\xi\sim \tau_q/\vo$ is controlled by the four-cycle $\tau_q$ which gets charged under the anomalous $U(1)$ and whose volume resolves the singularity. A leading order supersymmetric solution fixes $|\hat{C}|^2 = \xi/q$, leaving a flat direction in the $(|\hat{C}|,\tau_q)$-plane. This remaining flat direction is fixed by subdominant supersymmetry breaking contributions from background fluxes which take the form \cite{Cicoli:2013cha}:
\be
V_F(|\hat{C}|) = c_2 m_0^2 |\hat{C}|^2 + c_3 A |\hat{C}|^3 + O(|\hat{C}|^4)\,,
\label{VFC}
\ee
where $c_2$ and $c_3$ are $\mc{O}(1)$ coefficients. If we parametrise the volume dependence of the soft scalar masses as $m_0 \sim M_p/\vo^{\alpha_2}$ and the trilinear A-term as $A\sim M_p/\vo^{\alpha_3}$, and we use the vanishing D-term condition to write $\tau_q$ in terms of $|\hat{C}|$ as $\tau_q \sim |\hat{C}|^2 \vo$, the matter field VEV scales as:
\bea
(i)\,\,\text{If}\,\,c_2&>&0\quad |\hat{C}| = 0 \quad \Leftrightarrow\quad \tau_q=0\,,      \nn \\
(ii)\,\,\text{If}\,\,c_2&<&0\quad |\hat{C}| \simeq \frac{M_p}{\vo^{2\alpha_2-\alpha_3}} \quad \Leftrightarrow\quad  
\tau_q\simeq \frac{1}{\vo^{4\alpha_2-2\alpha_3-1}}\,. \nn
\eea
Only in case $(ii)$ the matter field $|\hat{C}|$ becomes tachyonic and breaks the Peccei-Quinn symmetry, leading to a viable axion realisation. In the presence of flavour D7-branes intersecting the D3-brane stack at the singularity, the soft terms are unsequestered and $\alpha_2=\alpha_3=1$ \cite{Conlon:2010ji}, giving $|\hat{C}| \sim M_p / \vo\sim m_{3/2}$ and $\tau_q \simeq \vo^{-1}\ll 1$ which ensures that $\tau_q$ is still in the singular regime. If the internal volume is of order $\vo\sim 10^8$, the large modulus $\tau_b$ is heavy enough to avoid the cosmological moduli problem: $m_{\tau_b}\sim 100$ TeV. In turn the gravitino mass, all soft terms and the axion decay constant $f_{a_\ALP} = |\hat{C}|$ are around $10^9$ GeV. Setting $\theta = a_\ALP/ f_{a_\ALP}$, the axion to photon coupling then takes the form:
\be
\frac{g^2}{32\pi^2}\,\frac{a_\ALP}{f_{a_\ALP}}\,F_{\mu\nu}\tilde{F}^{\mu\nu} \quad\Leftrightarrow\quad 
M  =  \frac{32\pi^2\,f_{a_\ALP}}{g^2} = \frac{32\pi^2\,f_{a_\ALP}}{g_s}\sim 10^{12}\,{\rm GeV}\,,
\label{D7fa}
\ee
since for D3-branes the coupling $g^{-2} = {\rm Re}(S) = g_s^{-1}$ is set by the dilaton $S$ which controls also the size of the string coupling that we assume to be in the perturbative regime: $g_s\simeq 0.1$.

On the other hand, in the absence of flavour D7-branes the soft terms are sequestered with $\alpha_3=2$ and $\alpha_2=3/2$ or $\alpha_2=2$ depending on the form of the quantum corrections to the K\"ahler metric for matter fields and the effects responsible for achieving a dS vacuum \cite{Blumenhagen:2009gk, Aparicio:2014wxa}. Notice that possible non-perturbative desequestering effects from couplings in the superpotential of the form $W_{\rm np}\supset  \mc{O}_{\rm matter}\, e^{-a_s T_s}$ with $\mc{O}_{\rm matter}$ a gauge-invariant operator composed of matter fields, cannot actually change the volume dependence of either the soft scalar masses or the A-terms \cite{Berg:2010ha}. Thus if $\alpha_2 = 3/2$ we have $f_{a_\ALP} = |\hat{C}| \simeq M_p/\vo$ and $\tau_q \sim \vo^{-1}\ll 1$, while if $\alpha_2 = 2$ the open axion decay constant scales as $f_{a_\ALP} = |\hat{C}| \simeq M_p/\vo^2$ and $\tau_q \sim \vo^{-3}\ll 1$. In both cases without flavour D7-branes the gaugino masses scale as $M_{1/2}\sim 0.1 \,M_p /\vo^2$ and lie around the TeV scale for $\vo \sim 10^7$. Considering this value of the volume, the axion-photon coupling therefore becomes:
\bea
(a)\,\,\text{If}\,\,\alpha_2&=&\frac32 \quad M  =  \frac{32\pi^2\,f_{a_\ALP}}{g_s} \sim 10^3 \,m_{3/2} \sim 10^{13} \,{\rm GeV} \,,  
\label{D3seq} \\
(b)\,\,\text{If}\,\,\alpha_2&=&2  \quad M  =  \frac{32\pi^2\,f_{a_\ALP}}{g_s} \sim 10^3 \,\frac{m_{3/2}}{\vo} \sim 10^6 \,{\rm GeV} \,. 
\label{D3superseq}
\eea

\item \textbf{ALP-photon coupling induced by $U(1)$ kinetic mixing}: We have shown above that, if the matter field $|\hat{C}|$ charged under the anomalous $U(1)$ develops a non-zero VEV due to a tachyonic soft scalar mass contribution, the open string axion $\theta$ can have an intermediate scale coupling to photons. However $\theta$ in general plays the r\^ole of the standard QCD axion which becomes much heavier than $m_\ALP\lesssim 10^{-12}$ eV due to QCD instanton effects. Hence the simplest realisation of an ultralight ALP with the desired phenomenological features to reproduce the $3.5$ keV line requires the existence of at least two open string axions. The Calabi-Yau volume (\ref{simplestLVS}) has then to be generalised to:
\be
\vo=\tau_b^{3/2}-\tau_s^{3/2}-\tau_{q_1}^{3/2}-\tau_{q_2}^{3/2}\,,
\label{simpleLVS}
\ee
where $\tau_{q_1}$ and $\tau_{q_2}$ are both collapsed to a singularity via D-term fixing and support a stack of D3-branes. There are two possibilities to realise a viable $a_\ALP$:
\begin{enumerate}
\item The two blow-up modes $\tau_{q_1}$ and $\tau_{q_2}$ are exchanged by the orientifold involution \cite{Cicoli:2012vw, Cicoli:2013mpa, Cicoli:2013cha}. The resulting quiver gauge theory on the visible sector stack of D3-branes generically features two anomalous $U(1)$ symmetries. This is for example always the case for del Pezzo singularities. Hence the visible sector is characterised by the presence of two open string axions: one behaves as the QCD axion while the other can be an almost massless $a_\ALP$ with $M\sim 10^{11}$-$10^{12}$ GeV as in (\ref{D7fa}) or (\ref{D3seq}). In this case the matter field $|\hat{C}|$ which develops a VEV of order the gravitino mass has to be a Standard Model gauge singlet in order not to break any visible sector gauge symmetry at a high scale.

\item The two blow-up divisors $\tau_{q_1}$ and $\tau_{q_2}$ are invariant under the orientifold involution \cite{Wijnholt:2007vn, Cicoli:2017shd}. Therefore one D3-stack has to reproduce the visible sector while the other represents a hidden sector. Each of the two sectors features an anomalous $U(1)$ which gives rise to an open string axion with a coupling to the respective photons controlled by the scale $M$. The visible sector axion plays the r\^ole of the QCD axion while the hidden sector open string axion can behave as $a_\ALP$. Its coupling to ordinary photons can be induced by a $U(1)$ kinetic mixing of the form \cite{Abel:2008ai, Goodsell:2009xc, Cicoli:2011yh}:
\be
\mc{L}\supset -\frac14 F_{\mu\nu} F^{\mu\nu} -\frac14 G_{\mu\nu} G^{\mu\nu} +\frac{\chi}{2} F_{\mu\nu} G^{\mu\nu} -\frac{a_\QCD}{4 M_{\rm vis}}\, F_{\mu\nu} \tilde{F}^{\mu\nu} -\frac{a_\ALP}{4 M_{\rm hid}}\, G_{\mu\nu} \tilde{G}^{\mu\nu}\,,
\label{Lmix}
\ee
where we denoted the QCD axion as $a_\QCD$, the kinetic mixing parameter as $\chi$ and the visible sector Maxwell tensor as $F_{\mu\nu}$ while the hidden one as $G_{\mu\nu}$. The kinetic mixing parameter is induced at one-loop level and scales as:
\be
\chi \sim \frac{g_{\rm vis} g_{\rm hid}}{16\pi^2}  = \frac{g_s}{16\pi^2} \simeq 10^{-3}\,.
\label{chi}
\ee
After diagonalising the gauge kinetic terms in (\ref{Lmix}) via $G_{\mu\nu} = G_{\mu\nu}' + \chi F_{\mu\nu}$, $a_\ALP$ acquires a coupling to ordinary photons of the form:
\be
\mc{L}\supset -\frac{\chi^2\, a_\ALP}{4 M_{\rm hid}} F_{\mu\nu} \tilde{F}^{\mu\nu}\qquad\Leftrightarrow\qquad M \simeq \frac{M_{\rm hid}}{\chi^2}\gg M_{\rm hid}\,.
\label{LmixNew}
\ee
Given that $M\gg M_{\rm hid}$, $a_\ALP$ can be a hidden sector open string axion only in case (\ref{D3superseq}) where the scale of the coupling to hidden photons of order $M_{\rm hid}\sim 10^6$ GeV is enhanced via $U(1)$ kinetic mixing to $M\sim 10^{12}$ GeV for the coupling to ordinary photons. 
\end{enumerate}

\item \textbf{DM as a local closed string axion fixed by poly-instanton effects:} In order to produce a monochromatic $3.5$ keV line, the DM mass has to be around $m_\DM\sim 7$ keV. Such a light DM particle can be a sterile neutrino realised as an open string mode belonging to either the visible or the hidden sector. However we shall focus on a more model-independent realisation of the decaying DM particle as a closed string axion. A generic feature of any 4D string model where the moduli are stabilised by perturbative effects, is the presence of very light axions whose mass is exponentially suppressed with respect to the gravitino mass \cite{Allahverdi:2014ppa}. Thus closed string axions are perfect candidates for ultra-light DM particles. In LVS models, there are two kinds of axions which remain light:
\begin{enumerate}
\item Bulk closed string axion $c_b$ since the corresponding supersymmetric partner $\tau_b$ is fixed by $\alpha'$ corrections to the K\"ahler potential $K$. This axionic mode develops a tiny mass only via $T_b$-dependent non-perturbative contributions to the superpotential $W$: $m_{c_b}\sim m_{\tau_b}\,e^{-\pi \tau_b}\ll m_{\tau_b} \sim m_{3/2}/\sqrt{\vo}$.

\item Local closed string axion $c_p$ whose associated modulus $\tau_p$ is stabilised by $g_s$ loop corrections to $K$. This can happen for so-called `Wilson divisors' $D_p$ which are rigid, i.e. $h^{2,0}(D_p)=0$, with a Wilson line, i.e. $h^{1,0}(D_p)=1$ \cite{Blumenhagen:2012kz}. Under these topological conditions, an ED3-instanton wrapping such a divisor does not lead to a standard non-perturbative contribution to $W$ but it generates a non-perturbative correction to another ED3-instanton wrapping a different rigid divisor $\tau_s$. This gives rise to poly-instanton corrections to $W$ of the form \cite{Blumenhagen:2008ji}:
\be
W_{\rm np} = A_s \,e^{-2\pi \left(T_s + A_p e^{- 2\pi T_p}\right) }\simeq A_s\,e^{-2\pi T_s}  - 2\pi A_s A_p\, e^{-2\pi T_s} e^{- 2\pi T_p} \,.
\label{Wpoly}
\ee
In LVS models, the blow-up mode $\tau_s$ is fixed by the dominant non-perturbative correction in (\ref{Wpoly}) since the leading loop contribution to the scalar potential is vanishing due to the `extended no-scale' structure \cite{Cicoli:2007xp}. Thus the corresponding axion $c_s$ becomes too heavy to play the r\^ole of $a_\DM$ since it acquires a mass of the same order of magnitude: $m_{\tau_s}\sim m_{c_s}\sim m_{3/2}$. On the other hand, the $T_p$-dependent non-perturbative correction in (\ref{Wpoly}) has a double exponential suppression, and so $\tau_p$ gets frozen by perturbative $g_s$ effects \cite{Berg:2005ja, Berg:2007wt}. Given that $c_p$ enjoys a shift symmetry which is broken only at non-perturbative level, this axion receives a potential only due to tiny poly-instanton contributions to $W$ which make it much lighter than $\tau_p$. Hence $c_p$ is a natural candidate for $a_\DM$ since $m_{c_p}\sim m_{\tau_p}\,e^{-\pi \tau_p/2} \ll m_{\tau_p}\sim m_{3/2}$. Notice that the presence of a `Wilson divisors' $\tau_p$ would modify the volume form (\ref{simpleLVS}) to \cite{Blumenhagen:2012kz}:
\be
\vo=\tau_b^{3/2}-\tau_s^{3/2} - (\tau_s+\tau_p)^{3/2}-\tau_{q_1}^{3/2}-\tau_{q_2}^{3/2}\,.
\label{voLVS}
\ee
\end{enumerate}

\item \textbf{DM to ALP decay induced by non-perturbative effects in $K$:} A DM to ALP coupling controlled by the scale $\Lambda$ of the form shown in (\ref{effL}) can arise from the kinetic mixing between a closed string DM axion and an open string ALP. Given that the kinetic terms are determined by the K\"ahler potential, a kinetic mixing effect can be induced by non-perturbative corrections to the K\"ahler metric for matter fields which we assume to take the form:\footnote{Similar non-perturbative corrections to $K$ induced by ED1-instantons wrapped around two-cycles have been computed for type I vacua in \cite{Camara:2008zk} and for type IIB vacua in \cite{RoblesLlana:2006is}, while similar non-perturbative effects in $K$ from an ED3-instanton wrapped around the K3 divisor in type I$'$ string theory, i.e. type IIB compactified on K3$\times T^2/\mathbb{Z}_2$, have been derived in \cite{Berglund:2005dm}.}
\be
K_{\rm np} \supset B_i\,e^{-b_i \tau_i}\,\cos(b_i c_i)\, C\bar{C}\,,
\ee
where $i=b$ if $a_\DM$ is a bulk closed string axion or $i=p$ if $a_\DM$ is a local closed string axion fixed by poly-instanton effects. As we shall show in Sec. \ref{DM-ALP} after performing a proper canonical normalisation of both axion fields, the resulting scale which controls the DM-ALP coupling is given by:
\bea
\Lambda\sim\left\lbrace
\begin{array}{l}
\displaystyle{
\frac{e^{b_b \tau_b}}{B_b\, \vo^{4/3}}}\,M_p \sim \displaystyle{
\frac{e^{b_b \vo^{2/3}}}{B_b\, \vo^{4/3}}}\,M_p \gg M_p \quad\text{for}\quad a_\DM= c_b \\ \\
\displaystyle{
\frac{e^{b_p \tau_p }}{B_p\, \vo^{7/6}}}\,M_p \sim \displaystyle{
\frac{M_p}{B_p\, \vo^{7/6-\kappa/N}}} \quad\text{for}\quad a_\DM= c_p 
\end{array}
\right.
\label{Lambdasize}
\eea
where $b_p=2\pi/N$, $\kappa = \tau_p/\tau_s$ and we have approximated $\vo \sim \tau_b^{3/2}\sim e^{2\pi\tau_s}$. From (\ref{Lambdasize}) it is clear that $\Lambda$ can be around the GUT/Planck scale only if the DM particle is a local closed string axion stabilised by tiny poly-instanton corrections to $W$ which can give it a small mass of order $m_\DM\sim 7$ keV.
\end{itemize}

\subsection{Calabi-Yau threefold}
\label{CY}

As explained in Sec. \ref{PhenoFeatures}, the minimal setup which can yield a viable microscopic realisation of the $a_\DM\to a_\ALP \to \gamma$ model for the $3.5$ keV line of \cite{Cicoli:2014bfa}, is characterised by a Calabi-Yau with $h^{1,1}=5$ K\"ahler moduli and a volume of the form (\ref{voLVS}). A concrete Calabi-Yau threefold built via toric geometry which reproduces the volume form (\ref{voLVS}) for $h^{1,1}=4$ (setting either $\tau_{q_1}=0$ or $\tau_{q_2}=0$) is given by example C of \cite{Blumenhagen:2012kz}. We therefore assume the existence of a Calabi-Yau threefold $X$ with one large divisor controlling the overall volume $D_b$, three del Pezzo surfaces, $D_s$, $D_{q_1}$ and $D_{q_2}$ and a `Wilson divisor' $D_p$. 

We expand the K\"ahler form $J$ in a basis of Poincar\'e dual two-forms as $J=t_b \hat{D}_b-t_s \hat{D}_s-t_{q_1} \hat{D}_{q_1}-t_{q_2} \hat{D}_{q_2}-t_p \hat{D}_p$, where the $t_i$'s are two-cycle volumes and we took a minus sign for the rigid divisors so that the corresponding $t_i$'s are positive. The Calabi-Yau volume then looks like:
\be
\vo=\frac16 \int_X J \wedge J \wedge J = \frac16 \left[ k_{bbb} t_b^3 - k_{sss} \left(t_s + \lambda t_p\right)^3 - \mu t_p^3 - k_{q_1 q_1 q_1} t_{q_1}^3 - k_{q_2 q_2 q_2} t_{q_2}^3 \right],
\label{Explvo}
\ee
where the coefficients $\lambda$ and $\mu$ are determined by the triple intersection numbers $k_{ijk}=\int_X \hat{D}_i\wedge\hat{D}_j\wedge\hat{D}_k$ as:
\be
\lambda = \frac{k_{ssp}}{k_{sss}} = \frac{k_{spp}}{k_{ssp}} \qquad\text{and}\qquad \mu = k_{ppp} - \frac{k_{ssp}^3}{k_{sss}^2} \,. \nn
\ee
The volume of the curve resulting from the intersection of the del Pezzo divisor $D_s$ with the Wilson surface $D_p$ is given by:
\be
{\rm Vol}(D_s\cap D_p) = \int_X J \wedge \hat{D}_s \wedge \hat{D}_p = -\left( k_{ssp} t_s + k_{spp} t_p \right) = - k_{ssp} \left( t_s + \lambda t_p \right)\,.
\ee
The volume of this curve is positive and the signature of the matrix $\frac{\partial^2\vo}{\partial t_i \partial t_j}$ is guaranteed to be $(1,h^{1,1}-1)$ (so with $1$ positive and $4$ negative eigenvalues) \cite{Candelas:1990pi} if $k_{ssp}<0$ while all the other intersection numbers are positive and $t_s + \lambda t_p >0$.\footnote{This analysis includes example C of \cite{Blumenhagen:2012kz} where $k_{sss}= k_{spp} = - k_{ssp} = 9$ and $k_{ppp} = 0$.}
The four-cycle moduli can be computed as:
\be
\tau_i=\frac12 \int_X J\wedge J\wedge \hat{D}_i\,,
\ee
and so they become:
\bea
\tau_b &=& \frac12 k_{bbb}\,t_b^2 \,,\qquad \tau_{q_1} =\frac12 k_{q_1 q_1 q_1}\,t_{q_1}^2\,,\qquad \tau_{q_2} =\frac12 k_{q_2 q_2 q_2}\,t_{q_2}^2\,, \nn \\
\tau_s &=& \frac12 \left(k_{sss}\,t_s^2+ k_{spp}\,t_p^2+2 k_{ssp}\,t_s t_p\right)=\frac12 k_{sss} \left( t_s + \lambda t_p \right)^2\,, \\
\tau_p &=& \frac12 \left(k_{ppp} \,t_p^2+ k_{ssp}\,t_s^2+2 k_{spp}\,t_s t_p\right)=\frac12 k_{ssp} \left( t_s + \lambda t_p \right)^2 +\frac12 \mu t_p^2\,. \nn
\eea
The overall volume (\ref{Explvo}) can therefore be rewritten in terms of the four-cycle moduli as:
\be
\vo= \lambda_b \tau_b^{3/2} - \lambda_s \tau^{3/2}_s -\lambda_p \left(\tau_p+ x \tau_s\right)^{3/2}-\lambda_{q_1} \tau_{q_1}^{3/2} - \lambda_{q_2} \tau_{q_2}^{3/2} \,,
\label{volume}
\ee
where:
\be
\lambda_i = \frac13 \sqrt{\frac{2}{k_{iii}}}\,,\quad \forall\,i=b,s,q_1,q_2\,,\qquad
\lambda_p = \frac13 \sqrt{\frac{2}{\mu}}\qquad\text{and}\qquad x = - \frac{k_{ssp}}{k_{sss}}>0\,. \nn
\ee
Notice that (\ref{volume}) reproduces exactly the volume form (\ref{voLVS}).

\subsection{Brane set-up and fluxes}
\label{setup}

As explained in Sec. \ref{PhenoFeatures}, $a_\ALP$ can be realised as an open string axion belonging either to the visible sector or to a hidden sector. In the first case the two rigid divisors $D_{q_1}$ and $D_{q_2}$ are exchanged by a proper orientifold involution whereas in the second case they are invariant. 
As we shall see more in detail in Sec. \ref{Dfix}, these two blow-up modes shrink down to zero size due to D-term stabilisation and support a stack of D3-branes at the resulting singularity.

Full moduli stabilisation requires the presence of non-perturbative corrections to the superpotential. We shall therefore consider an ED3-instanton wrapped around the `small' rigid divisor $D_s$ which generates a standard non-perturbative contribution to $W$, together with another ED3-instanton wrapped around the Wilson surface $D_p$ which gives rise to poly-instanton effects. In order to make $\tau_p$ heavier than the DM axion $c_p$, we need also to include a D7-stack that generates $\tau_p$-dependent string loop corrections to the K\"ahler potential. This can be achieved if a stack of D7-branes wraps the divisor $D_\D7$ (which we assume to be smooth and connected) given by:
\be
D_\D7=m_s \,D_s+m_p\,D_p\,, \quad \text{with}\quad m_s,m_p \in \mathbb{Z}\,.
\ee
In what follows we shall assume the existence of a suitable orientifold involution and O7-planes which allow the presence of such a D7-stack in a way compatible with D7-tadpole cancellation. The cancellation of Freed-Witten anomalies requires to turn on half-integer world-volume fluxes on the instantons and the D7-stack of the form \cite{Freed:1999vc}: 
\be
F_\D7 = f_s\,\hat{D}_s + f_p \,\hat{D}_p + \frac12\,\hat{D}_\D7\,,\qquad F_s = \frac12\,\hat{D}_s \,,
\qquad F_p = \frac12\,\hat{D}_p \,,
\ee
with $f_s$, $f_p \in \mathbb{Z}$. In order to guarantee a non-vanishing contribution to $W$, the total flux $\mc{F}_j = F_j -\iota^*_j B$ (with $\iota^*_j B$ the pull-back of the NSNS $B$-field on $D_j$) on both instantons has to be zero: $\mc{F}_s =\mc{F}_p=0$. This can be achieved if the $B$-field is chosen such that:
\be 
B = \frac12\,\hat{D}_s + \frac12\,\hat{D}_p\,,
\ee
and the pull-back of $\hat{D}_s/2$ on $D_p$ and of $\hat{D}_p/2$ on $D_s$ are both integer forms since in this case we can always turn on integer flux quanta to cancel their contribution to the total gauge flux. This is indeed the case if, for an arbitrary integer form $\omega = \omega_i \hat{D}_i\,\in\,H^2(\mathbb{Z},X)$ with $\omega_i \in \mathbb{Z}$, we have that:
\be
\frac12 \int_X \hat{D}_s \wedge \hat{D}_p \wedge \omega = \frac12 \left( k_{ssp} \,\omega_s +  k_{spp} \,\omega_p \right) \,\in\,\mathbb{Z}\,.
\ee
This condition can be easily satisfied if both $k_{ssp}$ and $k_{spp}$ are even. 

The total gauge flux on the D7-stack instead becomes:
\be
\mc{F}_\D7 = f_s\,\hat{D}_s + f_p \,\hat{D}_p + \frac12\left(m_s-1\right)\hat{D}_s + \frac12\left(m_p-1\right)\hat{D}_p = f_s\,\hat{D}_s + f_p \,\hat{D}_p\,, \nn
\ee
where without loss of generality, we have chosen $m_s=m_p=1$ so that $\mc{F}_\D7$ is an integer flux. The presence of this flux has several implications:
\begin{itemize}
\item The blow-up moduli $T_s$ and $T_p$ get charged under the diagonal $U(1)$ of the D7-stack with charges:
\be
q_i = \int_X \mc{F}_\D7 \wedge \hat{D}_\D7 \wedge \hat{D}_i = f_s \left(k_{ssi}+ k_{spi}\right) + f_p \left(k_{spi}+ k_{ppi}\right),\quad i=s,p\,,
\ee
which implies $q_p = \mu f_p - x\,q_s $.

\item The coupling constant of the gauge theory living on $D_\D7$ acquires a flux-dependent shift of the form:
\be
g_\D7^{-2}= \tau_s +\tau_p -h(\mc{F}_\D7)\, {\rm Re}(S)\,,
\label{gD7}
\ee
where ${\rm Re}(S)=e^{-\varphi}=g_s^{-1}$ is the real part of the axio-dilaton while the flux-dependent shift reads:
\be
h(\mc{F}_\D7) =\frac12 \int_X \mc{F}_\D7 \wedge \mc{F}_\D7 \wedge \hat{D}_\D7 = \frac{f_s}{2}\,q_s + \frac{f_p}{2}\,q_p\,. 
\label{hshift}
\ee

\item $\mc{F}_\D7$ generates a moduli-dependent FI-term which looks like:
\bea
\xi_\D7=\frac{1}{4\pi\,\vo}\int_X J \wedge \mc{F}_\D7 \wedge \hat{D}_\D7 = \frac{1}{4\pi\,\vo}\left(q_s\,t_s + q_p\,t_p\right).
\label{FID7}
\eea

\item A non-vanishing gauge flux on $D_\D7$ might induce chiral intersections between the D7 stack and the instantons on $D_s$ and $D_p$. Their net number is counted by the moduli $U(1)$-charges as:
\be
I_{\D7\text{-}\E3} = \int_X \mc{F}_\D7 \wedge \hat{D}_\D7 \wedge \hat{D}_s = q_s \,, \qquad
I_{\D7\text{-}{\rm poly}} = \int_X \mc{F}_\D7\wedge\hat{D}_\D7\wedge \hat{D}_p = q_p \,. 
\label{ChiralInt}
\ee
\end{itemize}

The relations (\ref{ChiralInt}) imply that, whenever an instanton has a non-vanishing chiral intersection with a stack of D-branes, the four-cycle modulus $T_{\rm inst}$ wrapped by the instanton gets charged under the diagonal $U(1)$ on the D-brane stack. Therefore a non-perturbative contribution to the superpotential of the form $W_{\rm np}\supset e^{-T_{\rm inst}}$ would not be gauge invariant. Thus a proper combination of $U(1)$-charged matter fields $\phi_i$ has to appear in the prefactor in order to make the whole contribution gauge invariant: $W_{\rm np}\supset \prod_i \phi_i \,e^{-T_{\rm inst}}$. If however the $\phi_i$ are visible sector matter fields, they have to develop a vanishing VEV in order not to break any Standard Model gauge group at high energies \cite{Blumenhagen:2007sm}. In our case the absence of chiral intersections between the instantons on $D_s$ and $D_p$ and the visible sector is guaranteed by the structure of the intersection numbers since $k_{s q_i j} = 0$ and $k_{p q_i j} = 0$ $\forall j$ for either $i=1$ or $i=2$. 

On the other hand, as can be seen from (\ref{ChiralInt}), there are chiral intersections between the hidden D7-stack on $D_\D7$ and the two instantons on $D_s$ and $D_p$. We could kill both of these intersections by setting $\mc{F}_\D7=0$. However this choice of the gauge flux on $D_\D7$ would also set to zero the FI-term in (\ref{FID7}) which is instead crucial to make $\tau_p$ heavier than the DM axion $c_p$. We shall therefore perform a choice of the gauge flux $\mc{F}_\D7$ which sets $I_{\D7\text{-}\E3} = q_s = 0$ but leaves $I_{\D7\text{-}{\rm poly}} = q_p \neq 0$ so that $\xi_\D7$ can develop a non-trivial dependence on $\tau_p$. This can take place if the flux quanta $f_p$ and $f_s$ are chosen such that:
\be
f_p = -\frac{k_{sss}+k_{ssp}}{k_{ssp} + k_{spp}}\,f_s\quad\Leftrightarrow\quad q_s=0\quad\text{and}\quad q_p = \mu f_p\,.
\ee
The FI-term in (\ref{FID7}) then becomes:
\bea
\xi_\D7= \frac{q_p}{4\pi}\,\frac{t_p}{\vo} =  \frac{f_p\sqrt{2\,\mu}}{4\pi}\, \frac{\sqrt{\tau_p + x\tau_s}}{\vo}\,,
\label{FIterm}
\eea
while the shift of the gauge coupling in (\ref{hshift}) simplifies to $h(\mc{F}_\D7) = \frac{\mu}{2}\,f_p^2$. Due to non-zero chiral intersections between the D7-stack and the divisor $D_p$, the poly-instanton contribution to the superpotential comes with a prefactor that depends on a $U(1)$-charged matter field $\phi$. In Sec. \ref{Dfix} we will show that the interplay between D-terms and string loop effects can fix $\phi$ at a non-zero VEV, so that the poly-instanton correction is non-vanishing. Notice that $\phi$ belongs to a hidden sector, and so it can safely develop a non-zero VEV at high energies without violating any phenomenological requirement.

\subsection{Low-energy 4D theory}
\label{4Dtheory}

Type IIB string theory compactified on an orientifold of the Calabi-Yau threefold described in Sec. \ref{CY} with the brane setup and gauge fluxes of Sec. \ref{setup} gives rise to an $N=1$ 4D supergravity effective field theory characterised by a K\"ahler potential $K$ and a superpotential $W$ of the form: 
\be
K = K_{\rm mod} + K_{\rm matter}\,\qquad\text{and}\qquad W = W_{\rm tree} + W_{\rm np}\,,
\ee
where:
\begin{itemize}
\item The moduli K\"ahler potential receives perturbative $\alpha'$ and $g_s$ corrections beyond the tree-level approximation:
\be
K_{\rm mod} = K_{\rm tree} + K_{\alpha'} + K_{g_s}\,,
\ee
with:
\be
K_{\rm tree} = -2 \ln \vo + \frac{\tau_{q_1}^2}{\vo} + \frac{\tau_{q_2}^2}{\vo}\,,\qquad
K_{\alpha'} = -\frac{\zeta}{g_s^{3/2} \vo}\,,\qquad K_{g_s} = g_s \sum_i \frac{C_i^\KK\,t_i^\perp}{\vo}\,.
\label{K}
\ee
In (\ref{K}) we neglected the tree-level K\"ahler potential for the dilaton $S = e^{-\varphi} - {\rm i}\,C_0$ and the complex structure moduli $U_{a_-}$, $a_-=1,\cdots,h^{1,2}_-$ and we expanded the effective theory around the singularities obtained by collapsing the two blow-up modes $\tau_{q_1}$ and $\tau_{q_2}$ (hence the volume $\vo$ in (\ref{K}) should be thought of as (\ref{volume}) with $\tau_{q_1}=\tau_{q_2} = 0$). Moreover, we included only the leading order $\alpha'$ correction which depends on $\zeta =-\frac{\zeta(3)\chi(X)}{2(2\pi)^3}$ \cite{Becker:2002nn} since in the large volume limit higher derivative $\alpha'$ effects yield just subdominant contributions \cite{Ciupke:2015msa}. Finally in $K_{g_s}$ we considered only string loop corrections arising from the exchange of Kaluza-Klein modes between non-intersecting stacks of D-branes and O-planes ($C_i^\KK$ are complex structure dependent coefficients and $t_i^\perp$ is the two-cycle controlling the distance between two parallel stacks of D-branes/O-planes) while we did not introduce any $g_s$ effects coming from the exchange of winding modes since these arise only in the presence of intersections between D-branes which are however absent in our setup \cite{Cicoli:2007xp,Berg:2005ja,Berg:2007wt}. 

\item In the matter K\"ahler potential we focus just on the dependence on the matter fields which will develop a non-zero VEV. These are two $U(1)$-charged matter fields: $\phi = |\phi|\,e^{{\rm i}\psi}$ which belongs to the hidden D7-stack on $D_\D7$ and $C = |C|\,e^{{\rm i}\theta}$ which can be either a visible sector gauge singlet (if $D_{q_1}$ and $D_{q_2}$ are exchanged by the orientifold involution) or a hidden sector field (if both $D_{q_1}$ and $D_{q_2}$ are invariant under the orientifold involution) living on a D3-brane stack \cite{Aparicio:2008wh,Conlon:2006tj}:
\be
K_{\rm matter} = \frac{\phi\bar{\phi}}{{\rm Re}(S)} + \tilde{K} (T_i,\bar{T}_i)\,C\bar{C}\,.
\label{Kmatter}
\ee
In (\ref{Kmatter}) we wrote down just the tree-level K\"ahler metric for $\phi$ while we shall consider both perturbative and non-perturbative corrections to the K\"ahler metric for $C$ which we assume to take the form:
\be
\tilde {K}(T_i,\bar{T}_i)=\frac{f(S,U)}{\vo^{2/3}} + \tilde{K}_{\rm pert} + B_i \,e^{-b_i\tau_i}\,\cos(b_i c_i)\qquad\text{with}\quad i=b,p\,,
\label{Ktilde}
\ee
where $f(S,U)$ is an undetermined function of the dilaton and complex structure moduli, $\tilde{K}_{\rm pert}$ represents perturbative corrections which do not depend on the axionic fields because of their shift symmetry and the last term is a non-perturbative correction which can in principle depend on either the large or the poly-instanton cycle. This term induces a kinetic mixing between the open string axion $\theta$ and either of the two ultra-light closed string axions $c_b$ and $c_p$. As we shall see in Sec. \ref{Dfix}, the open string axion $\psi$ gets eaten up by the anomalous $U(1)$ on the D7-stack, and so light closed string axions cannot decay to this heavy mode. This is the reason why we did not include any non-perturbative effect in the K\"ahler metric for $\phi$. 

\item The tree-level superpotential $W_{\rm tree} = \int_X G_3 \wedge \Omega$, with $\Omega$ the Calabi-Yau $(3,0)$-form, is generated by turning on background three-form fluxes $G_3=F_3-{\rm i}\, S H_3$ and depends just on the dilaton and the $U$-moduli but not on the $T$-moduli \cite{Giddings:2001yu}.

\item The non-perturbative superpotential receives a single contribution from the ED3-instanton wrapped around $D_s$ together with poly-instanton effects from the ED3-instanton wrapped around the Wilson surface $D_p$ and takes the same form as (\ref{Wpoly}):
\be
W_{\rm np} = A_s\,e^{-2\pi T_s}  -2\pi A_s A_p\, e^{-2\pi T_s} e^{- 2\pi T_p} \,.
\label{Wpoly2}
\ee
The prefactors $A_s$ and $A_p$ depend on $S$ and $U$-moduli. Given that $T_p$ is charged under the anomalous diagonal $U(1)$ on the D7-stack, $A_p$ has to depend also on the charged matter field $\phi$ in order to make $W_{\rm np}$ gauge invariant. If we make the dependence of $A_p$ on $\phi$ explicit by replacing $A_p \to A_p \phi^n$ with arbitrary $n$, and we use the fact that $\phi$ and $T_p$ behave under a $U(1)$ transformation as:
\be
\delta \phi = {\rm i}\, q_\phi\,\phi \qquad\text{and}\qquad \delta T_p= {\rm i}\, \frac{q_p}{2\pi}\,,
\ee
the variation of $W_{\rm np}$ under a $U(1)$ transformation becomes:
\be
\delta W_{\rm np}= W_{\rm np}\left(n\, \frac{\delta\phi}{\phi} - 2\pi\delta T_p \right) = {\rm i}\, W_{\rm np}\left(n\, q_\phi - q_p \right).
\ee
Hence $W$ is gauge invariant only if $n = q_p/q_\phi$. Notice that $n>0$ since, as we shall see in Sec. \ref{Dfix}, a consistent D-term stabilisation can yield a non-zero VEV for $\phi$ only if $q_\phi$ and $q_p$ have the same sign.
\end{itemize}

\section{Moduli stabilisation}
\label{ModStab} 

In this section we shall show how to stabilise all closed string moduli together with the two charged open string modes $\phi$ and $C$. The total $N=1$ supergravity scalar potential descending from the $K$ and $W$ described in Sec. \ref{4Dtheory}, includes both F- and D-term contributions of the form:
\be
V = V_F + V_D = e^K \left(K^{I\bar{J}} D_I W D_{\bar{J}}\bar{W}-3|W|^2\right) 
+ \frac{g_\D7^2}{2}\,D_\D7 ^2 + \frac{g_{\scriptscriptstyle D3}^2}{2}\,D_{\scriptscriptstyle D3}^2\,,
\label{Vfull}
\ee
where the K\"ahler covariant derivative is $D_I W= \partial_I W+ W \partial_I K$, the gauge coupling of the field theory living on the D7-stack is given by (\ref{gD7}) while $g_{\scriptscriptstyle D3}^{-2}={\rm Re}(S)$ for the quiver gauge theory on the D3-stack. The two D-term contributions look like:
\be
D_\D7 = q_\phi \,\phi\,\frac{\partial K}{\partial \phi} -\xi_\D7\,, \qquad \text{and}\qquad
D_{\scriptscriptstyle D3} = q_{\scriptscriptstyle C} \,C\,\frac{\partial K}{\partial C} -\xi_{\scriptscriptstyle D3}\,,
\ee
where the FI-term for the D7-stack is given by (\ref{FIterm}) whereas the FI-term for the D3-brane stack is:
\be
\xi_{\scriptscriptstyle D3}= q_i\, \frac{\partial K}{\partial T_{q_i}} = q_i\,\frac{\tau_{q_i}}{\vo}\qquad\text{for either}\,\,i=1\,\,\text{or}\,\,i=2\,.
\label{FI}
\ee
In LVS models the Calabi-Yau volume is exponentially large in string units, and so $1/\vo\ll 1$ is a small parameter which can be used to control the relative strength of different contributions to the total scalar potential (\ref{Vfull}). Let us analyse each of these contributions separately.

\subsection{Stabilisation at $\mc{O}(1/\vo^2)$}
\label{Dfix}

As can be seen from the volume scaling of the two FI-terms (\ref{FIterm}) and (\ref{FI}), the total D-term potential scales as $V_D \sim M_p^4/\vo^2 \sim M_s^4$. Therefore its leading order contribution has to be vanishing since otherwise the effective field theory would not be under control since the scalar potential would be of order the string scale. As we shall see in more detail below, this leading order supersymmetric stabilisation fixes $|\phi|$ in terms of $\tilde{\tau}_p \equiv \tau_p+x\tau_s$ and $\tau_{q_i}$ in terms of $|C|$. The open string axion $\psi$ and the closed string axion $c_{q_i}$ are eaten up by the two anomalous $U(1)$'s living respectively on the D7 and D3-stack. Additional $\mc{O}(1/\vo^2)$ tree-level contributions to the scalar potential arise from background fluxes which stabilise the dilaton and the complex structure moduli in a supersymmetric manner at $D_S W_{\rm tree} = D_{U_{a_-}} W_{\rm tree}=0$ \cite{Giddings:2001yu}. At this level of approximation the K\"ahler moduli are still flat due to the no-scale cancellation. They can be lifted by subdominant corrections to the effective action which can be studied by assuming a constant tree-level superpotential $W_0 =\langle W_{\rm tree}\rangle$ that is naturally of $\mc{O}(1)$. Summarising the total $\mc{O}(1/\vo^2)$ contribution to the scalar potential looks schematically like (we show the dependence just on the scalar fields which get frozen):
\be
V_{\mc{O}(1/\vo^2)} = V_D (|\phi|, \tau_{q_i}) + V_F^{\rm tree} (S,U)\,.
\ee
Let us focus in particular on the dynamics of the total D-term potential which from (\ref{FIterm}), (\ref{Vfull}) and (\ref{FI}) reads:
\be
V_D = \frac{g_\D7^2}{2}\left(q_\phi \,\frac{|\phi|^2}{{\rm Re}(S)} -\frac{f_p\sqrt{2\,\mu}}{4\pi}\, \frac{\sqrt{\tilde{\tau}_p}}{\vo}\right) ^2 
+ \frac{g_{\scriptscriptstyle D3}^2}{2}\,\left(q_{\scriptscriptstyle C} \,\tilde{K} (T_i,\bar{T}_i)\,|C|^2 -q_i\,\frac{\tau_{q_i}}{\vo}\right)^2\,.
\ee
Supersymmetry is preserved if: 
\be
q_\phi \,\frac{|\phi|^2}{{\rm Re}(S)} = \frac{f_p\sqrt{2\,\mu}}{4\pi}\, \frac{\sqrt{\tilde{\tau}_p}}{\vo}\qquad\text{and}\qquad
q_{\scriptscriptstyle C} \,\tilde{K} (T_i,\bar{T}_i)\,|C|^2 = q_i\,\frac{\tau_{q_i}}{\vo}\,.
\label{Dstab}
\ee
These two relations fix one direction in the $(|\phi|, \tilde{\tau}_p)$-plane and one direction in the $(|C|,\tau_{q_i})$-plane. Each of these two directions corresponds to the supersymmetric partner of the axion which is eaten up by the relative anomalous $U(1)$ gauge boson in the process of anomaly cancellation. 
The axions which become the longitudinal components of the massive gauge bosons are combinations of open string axions with decay constant $f_{\rm op}$ and closed string axions with decay constant $f_{\rm cl}$. The St\"uckelberg mass of the anomalous $U(1)$'s scales as \cite{ArkaniHamed:1998nu}: 
\be
M^2_{U(1)} \simeq g^2 \left( f_{\rm op}^2 + f_{\rm cl}^2 \right) \,,
\ee 
where:
\bea
\text{D7 case:} \quad f_{\rm op}^2 &=& \frac{|\phi|^2}{{\rm Re}(S)} = \frac{f_p\sqrt{2\mu}}{4\pi\,q_\phi}\, \frac{\sqrt{\tilde{\tau}_p}}{\vo} \gg f_{\rm cl}^2 = \frac14 \frac{\partial^2 K}{\partial \tau_p^2} =  \frac{1}{4\sqrt{2\mu}}\frac{1}{\vo\sqrt{\tilde{\tau}_p}}\,, \nn \\ 
\text{D3 case:} \quad f_{\rm op}^2 &=& \tilde{K} (T_i,\bar{T}_i)\,|C|^2 = \frac{q_i}{q_{\scriptscriptstyle C}}\,\frac{\tau_{q_i}}{\vo} \ll  f_{\rm cl}^2 = \frac14 \frac{\partial^2 K}{\partial \tau_{q_i}^2} = \frac{1}{2\vo}\,,
\label{Stuckel}
\eea
for:
\be
\tilde{\tau}_p\gg z_p \equiv \frac{\pi\,q_\phi}{2\mu f_p} \qquad\text{and}\qquad \tau_{q_i} \ll z_{q_i} \equiv \frac{q_{\scriptscriptstyle C}}{2q_i} \,. 
\label{Stuck}
\ee
In Sec. \ref{Ffix} and \ref{Ffix2} we shall explain how to fix the remaining flat directions, showing that the conditions in (\ref{Stuck}) can be satisfied dynamically. These conditions imply that for the D7 case the combination of axions eaten up is mostly given by the open string axion $\psi$, and so (\ref{Dstab}) should be read off as fixing $|\phi|$ in terms of $\tilde{\tau}_p$, while for the D3 brane case the combination of axions eaten up is mostly given by the closed string axion $c_{q_i}$ which means that (\ref{Dstab}) fixes $\tau_{q_i}$ in terms of $|C|$. Notice that from (\ref{Stuck}) the $U(1)$ gauge bosons acquire a mass of order the string scale: $M_{U(1)} \sim M_p /\sqrt{\vo}\sim M_s$.

\subsection{Stabilisation at $\mc{O}(1/\vo^3)$}
\label{Ffix}

As we shall explain more in detail below, $\mc{O}(1/\vo^3)$ effects arise from both the leading $\alpha'$ and $\tilde{\tau}_p$-dependent $g_s$ corrections to $K$ in (\ref{K}) together with the single instanton contribution in (\ref{Wpoly2}). They give rise to a scalar potential which depends on $\tau_s$, $c_s$, $\tau_p$ and $\tau_b$ but not on the associated axions $c_p$ and $c_b$ since both $T_p$- and $T_b$-dependent non-perturbative corrections to $W$ are much more suppressed due to the double exponential suppression of poly-instanton effects and the exponentially large value of $\tau_b\sim \vo^{2/3}$. These $\mc{O}(1/\vo^3)$ contributions alone would yield an AdS minimum which breaks supersymmetry spontaneously \cite{Balasubramanian:2005zx, Conlon:2005ki, Cicoli:2008va}. Additional contributions of the same order of magnitude can arise rather naturally from a hidden D7 T-brane stack \cite{Cicoli:2015ylx} or from anti-D3 branes at the tip of a warped throat \cite{Kachru:2003aw, Bergshoeff:2015jxa,Garcia-Etxebarria:2015lif} and can be tuned to obtain a dS vacuum. The K\"ahler moduli develop non-zero F-terms and mediate supersymmetry breaking to each open string sector via gravitational interactions. Matter fields on the D7-stack are unsequestered, and so acquire soft masses of order $m_{3/2}$. After using the vanishing D-term condition to write $|\phi|$ in terms of $\tilde{\tau}_p$, the resulting F-term potential for the matter fields also scales as $\mc{O}(1/\vo^3)$. Thus the full $\mc{O}(1/\vo^3)$ scalar potential behaves as:
\be
V_{\mc{O}(1/\vo^3)} = V_F^{\alpha'} (\vo) + V_F^{g_s} (\vo, \tilde{\tau}_p) + V_F^\E3 (\tau_s, c_s, \vo) + V_F^{\rm matter} (\vo, \tilde{\tau}_p) +  V_{\rm up}(\vo) \,.
\ee
All these $\mc{O}(1/\vo^3)$ contributions take the following precise form:
\bea
&& V_F^{\alpha'} (\vo) = \frac{3\,\zeta}{32\pi\,\sqrt{g_s}}\frac{W_0^2}{\vo^3}\,, \qquad 
V_F^{g_s} (\vo, \tilde{\tau}_p) = \frac{3\,g_s\,\lambda_p}{64\pi}\left(g_s\,C_p^\KK\right)^2\frac{W_0^2}{\vo^3\sqrt{\tilde{\tau}_p}}\,, \nn \\
&& V_F^\E3 (\tau_s, c_s, \vo) = \frac{4 g_s \pi A_s^2}{3\lambda_s}\frac{\sqrt{\tau_s}\,e^{-4\pi\tau_s}}{\vo}
+  g_s A_s \cos(2\pi c_s) \,\frac{W_0\,\tau_s\,e^{-2\pi\tau_s}}{\vo^2}\,,  \nn \\
&& V_F^{\rm matter} (\vo, \tilde{\tau}_p) = m_{3/2}^2 \frac{|\phi|^2}{{\rm Re}(S)} 
= \frac{3\,g_s\,\lambda_p}{64\pi z_p}\, \frac{W_0^2 \sqrt{\tilde{\tau}_p}}{\vo^3}\,, 
\label{VO3}
\eea
where the string loop potential includes only the leading Kaluza-Klein contribution from $K_{g_s}$ in (\ref{K}) which is given by \cite{Cicoli:2007xp}:
\be 
V_F^{g_s}(\vo, \tilde{\tau}_p) = \left(\frac{g_s}{8\pi}\right)\left(g_s\,C_p^\KK\right)^2  \frac{W_0^2}{\vo^2} \frac{\partial^2 K}{\partial\tau_p^2}\,, \nn
\ee  
and in $V_F^{\rm matter}$ we substituted the relation (\ref{Dstab}) which expresses $|\phi|$ in terms of $\tilde{\tau}_p$. Summing up the four contributions in (\ref{VO3}), the total scalar potential at $\mc{O}(1/\vo^3)$ has a minimum at (for $2\pi\tau_s \gg 1$):
\bea
\label{LVSmin}
c_s &=& k+\frac12 \quad \text{with}\,\,k\in\mathbb{Z}\,,\qquad \vo=\frac{3 \lambda_s}{8 \pi A_s} \,W_0 \sqrt{\tau_s} \,e^{2\pi\tau_s}\,,  \\
\tau_s &=& \left(\frac{\zeta}{2\lambda_s}\right)^{2/3} \frac{1}{g_s} \left(1 + \epsilon\right)\sim \frac{1}{g_s}\,,
\qquad \tilde{\tau}_p = z_p\,(g_s\,C_p^\KK)^2\sim \frac{1}{g_s}\,, \nn
\eea
for $C_p^\KK \sim g_s^{-3/2} \gg 1$ and: 
\be
\epsilon = \left(\frac{2\lambda_p}{3 \zeta z_p}\right) g_s^{3/2}\,\sqrt{ \tilde{\tau}_p} \sim g_s^{5/2}\,C_p^\KK \sim g_s \ll 1\,.
\label{eps}
\ee
Notice that the condition $\tilde{\tau}_p \gg z_p$ in (\ref{Stuck}), which ensures that the closed string axion $c_p$ is not eaten up by the anomalous $U(1)$ on the D7-stack and so can play the r\^ole of DM, can be easily satisfied if $C_p^\KK \sim g_s^{-3/2} \gg 1$. We point out that the coefficients of the string loop corrections are complex structure moduli dependent, and so their values can be tuned by appropriate choices of background fluxes. Therefore for $z_p\sim\mc{O}(1)$, $\tilde{\tau}_p\sim \tau_s \sim \tau_p \sim g_s^{-1} \gg 1$. This behaviour justifies also the scaling of the small parameter $\epsilon$ in (\ref{eps}).

As stressed above, this minimum is AdS but can be uplifted to dS via several different positive definite contributions. Two examples which emerge rather naturally in type IIB flux compactifications are T-branes \cite{Cicoli:2015ylx} or anti-D3 branes \cite{Kachru:2003aw, Bergshoeff:2015jxa,Garcia-Etxebarria:2015lif}.

\subsection{Stabilisation at $\mc{O}(1/\vo^{3+p})$}
\label{Ffix2}

Taking into account all contributions to the scalar potential up to $\mc{O}(1/\vo^3)$, there are still four flat directions: the charged matter field $|C|$, the open string axion $\theta$ and the two closed string axions $c_p$ and $c_b$. We shall now show how to stabilise the DM axion $c_p$ and $|C|$ which sets the decay constant of the ALP $\theta$ and fixes $\tau_{q_i}$ from (\ref{Dstab}). The bulk closed string axion $c_b$ receives scalar potential contributions only from $T_b$-dependent non-perturbative corrections, and so it is almost massless: $m_{c_b}\sim m_{\tau_b}\,e^{-\pi\,\vo^{2/3}} \sim 0$. 

The closed string axion $c_p$ and the open string matter field $|C|$ receive a potential respectively via poly-instanton corrections to the effective action and soft supersymmetry breaking terms. As we shall see below, these terms scale as $\mc{O}(1/\vo^{3+p})$ with $p>0$. The only exception which leads to $p=0$ is the case where flavour D7-branes desequester the open string sector on the D3-brane at a singularity. However, as shown in Sec. \ref{PhenoFeatures}, these effects would not modify the VEV of $|C|$ which sets the open string axion decay constant, and so, without loss of generality, we shall consider just the sequestered case. The resulting $\mc{O}(1/\vo^{3+p})$ scalar potential looks schematically as (showing again just the dependence on the fields which get stabilised at this order in the inverse volume expansion of $V$):
\be
V_{\mc{O}(1/\vo^{3+p})} = V_F^{\rm poly} (c_s) + V_F^{\rm matter} (|C|) \,.
\ee
The leading order expression of the $C$-dependent soft supersymmetry breaking terms is given by (\ref{VFC}). A more complete expression in terms of the canonically normalised field $\hat{C} = |\hat{C}|\,e^{{\rm i} \theta} = \sqrt{\tilde{K}}\,C$ (see Sec. \ref{CanNorm} for more details) is (the $c_i$'s are $\mc{O}(1)$ coefficients) \cite{Cicoli:2013cha}: 
\be
V_F(|\hat{C}|) = c_2 m_0^2 |\hat{C}|^2 + c_3 A |\hat{C}|^3 + c_4 \lambda |\hat{C}|^4 + \mc{O} (|\hat{C}|^5) + c_5 \frac{\tau_{q_i}^2}{\vo^3}\left[1+\mc{O}\left(\frac{1}{\vo}\right)\right] \,,
\label{VFC2}
\ee
where the first three terms originate from expanding the F-term potential in powers of $|\hat{C}|$ up to fourth order, whereas the last term comes from the fact that the $\tau_{q_i}$-dependent term in (\ref{K}) breaks the no-scale structure. Using (\ref{Dstab}) we can rewrite the last term in (\ref{VFC2}) in terms of $|\hat{C}|$ and parameterising the soft terms in Planck units as $m_0\sim \vo^{-\alpha_2}$, $A\sim \vo^{-\alpha_3}$ and $\lambda\sim \vo^{-\alpha_4}$, we obtain (up to fourth order in $|\hat{C}|$):
\be
V_F(|\hat{C}|) = \frac{c_2}{\vo^{\alpha_2}}\, |\hat{C}|^2 + \frac{c_3}{\vo^{\alpha_3}} |\hat{C}|^3 + \frac{k_4}{\vo^{\alpha_4}} \,|\hat{C}|^4
\qquad\text{with}\qquad k_4 = c_4 \lambda + \frac{4 c_5 z_{q_i}^2}{\vo^{1-\alpha_4}}\,.
\label{VFC3}
\ee
If the soft masses are non-tachyonic, the VEV of the matter field $|\hat{C}|$ is zero, and so the open string axion $\theta$ cannot play the r\^ole of the ALP $a_\ALP$ which gives the $3.5$ keV line by converting into photons in astrophysical magnetic fields. On the other hand, as explained in Sec. \ref{PhenoFeatures}, if $c_2<0$ $|\hat{C}|$ can develop a non-vanishing VEV. Open string modes living on D3-branes localised at singularities are geometrically sequestered from the sources of supersymmetry breaking in the bulk, resulting in $\alpha_3=2$, $\alpha_4=1$ and $\alpha_2=3/2$ or $\alpha_2=2$ depending on the exact moduli dependence of $\tilde{K}_{\rm pert}$ in (\ref{Ktilde}) and the details of the uplifting mechanism to a dS vacuum \cite{Blumenhagen:2009gk, Aparicio:2014wxa}. The VEVs of $|\hat{C}|$ and $\tau_{q_i}$ from (\ref{Dstab}) are therefore:
\bea
\alpha_2 = \frac32 \,\,\text{case:}\qquad |\hat{C}| &=& f_{a_\ALP} = \frac{M_p}{\vo}\,\,\,\qquad\Leftrightarrow\qquad \tau_{q_i} = \frac{2\,z_{q_i}}{\vo}\ll z_{q_i}\,, \label{fa1} \\
\alpha_2 = 2 \,\,\text{case:}\qquad |\hat{C}| &=& f_{a_\ALP} = \frac{M_p}{\vo^2}\qquad\Leftrightarrow\qquad \tau_{q_i} = \frac{2\,z_{q_i}}{\vo^3}\ll  z_{q_i}\,, \label{fa2}
\eea
where we have identified the open string axion $\theta$ with the ALP $a_\ALP = f_{a_\ALP}\,\theta$. Notice that the ALP decay constant in (\ref{fa1}) reproduces exactly the ALP coupling to gauge bosons in (\ref{D3seq}) while the $f_{a_\ALP}$ in (\ref{fa2}) gives the coupling in (\ref{D3superseq}). 
We stress that (\ref{fa1}) and (\ref{fa2}) show also how the condition $\tau_{q_i} \ll z_{q_i}$ in (\ref{Stuck}) is easily satisfied for $1/\vo \ll 1$. This ensures that the blow-up mode $\tau_{q_i}$ is indeed collapsed to a singularity. Let us remind the reader that $i$ can be either $i=2$ or $i=3$. When $\tau_{q_1}$ and $\tau_{q_2}$ are identified by the orientifold involution, an open string axion is the standard QCD axion $a_\QCD$ while the other is $a_\ALP$ with $|\hat{C}|$ a Standard Model gauge singlet with a large VEV. On the other hand, when the two blow-up modes $\tau_{q_1}$ and $\tau_{q_2}$ are separately invariant under the involution, $\hat{C}$ belongs to a hidden sector and, as described in Sec. \ref{PhenoFeatures}, its axion $\theta$ has a coupling to ordinary photons of the form (\ref{LmixNew}) which is induced by $U(1)$ kinetic mixing.

The axionic partner $c_p$ of the K\"ahler modulus $\tau_p$ which controls the volume of the Wilson divisor supporting poly-instanton effects, receives the following scalar potential contributions from the second term in (\ref{Wpoly2}) with $A_p\to A_p \phi^n$ and $n = q_p/q_\phi$:
\bea
V_F^{\rm poly}(c_p)&=&-2 g_s \pi A_s A_p \phi^n \left[\frac{8 (1-x) \pi A_s}{3\lambda_s}\,\cos(2\pi c_p)\,
\sqrt{\tau_s}\,e^{-2\pi\tau_s}\right. \nn \\
 &+& \left. W_0\,\cos[2\pi (c_s + c_p) ]\,\frac{\left((1-x)\tau_s+ \tilde{\tau}_p\right)}{\vo}\right]
\frac{e^{-2\pi\tau_s}\,e^{-2\pi \tau_p}}{\vo}\,, \nn
\eea
which, after using the first D-term relation in (\ref{Dstab}) and substituting the VEVs in (\ref{LVSmin}), reduces to (setting without loss of generality $\phi=|\phi|$ with $\psi=0$):
\be
V_F^{\rm poly} (c_p) = \frac{A}{\vo^{3+p}}\,\cos(2\pi c_p)\,,
\label{Vpoly}
\ee
where:
\be
A = \frac{3 g_s \lambda_s A_p}{4} \left(\frac{3\lambda_p\,C_p^\KK}{8\sqrt{z_p}} \right)^{n/2} \left(\frac{3 \lambda_s\sqrt{\tau_s}}{8 \pi A_s}\right)^\kappa \tilde{\tau}_p\,\sqrt{\tau_s} \,W_0^{2+\kappa}\,, \nn
\ee
with:
\be
\kappa \equiv \frac{\tau_p}{\tau_s}>0 \qquad\text{and}\qquad p = \frac{n}{2} + \kappa > 0\,.
\ee
Therefore the DM axion $c_p$ is stabilised at $\mc{O}(1/\vo^{3+p})$ at $c_p=1/2 +k$ with $k\in\mathbb{Z}$ and $A>0$.

\section{Mass spectrum and couplings}
\label{MassCoupl}

In this section we shall first determine the expressions for all canonically normalised fields and their mass spectrum, and then we will compute the strength of the coupling of the light DM axion $c_p$ to the open string ALP $\theta$ which is induced by non-perturbative corrections to the matter K\"ahler metric in (\ref{Ktilde}).

\subsection{Canonical normalisation}
\label{CanNorm}

Similarly to the scalar potential, also the kinetic Lagrangian derived from the K\"ahler potential for the moduli given by the three terms in (\ref{K}) and for the matter fields given by (\ref{Kmatter}), can be organised in an expansion in $1/\vo\ll 1$. Hence the kinetic terms can be canonically normalised order by order in this inverse volume expansion. The detailed calculation is presented in App. \ref{AppCanNorm} and here we just quote the main results which are useful to work out the strength of the DM-ALP coupling. The expressions for the canonically normalised fields at leading order look like (the moduli and the matter fields are dimensionless while canonically normalised scalar fields have standard mass dimensions):
\bea
\frac{|\hat{C}|}{M_p} &=& \sqrt{2\tilde{K}} |C| \,,\qquad a_\ALP = |\hat{C}| \theta  = f_{a_\ALP} \theta\,, \qquad
\frac{|\hat\phi|}{M_p} = \sqrt{\frac{2}{{\rm Re}(S)}}\, |\phi| \,,\qquad \frac{\phi_{q_i}}{M_p} = \frac{\tau_{q_i}}{\sqrt{\vo}}\,, \nn \\
\frac{\phi_b}{M_p} &=& \sqrt{\frac32}\,\ln\tau_b \,,\qquad \frac{a_b}{M_p} = \sqrt{\frac32} \frac{c_b}{\tau_b}\,, \qquad
\frac{\phi_s}{M_p} = \sqrt{\frac{4 \lambda_s}{3\vo}}\,\tau_s^{3/4}\,, \label{CN} \\
\frac{a_s}{M_p} &=& \sqrt{\frac{3 \lambda_s}{4\vo\sqrt{\tau_s}}}\,c_s\,,   \qquad
\frac{\tilde{\phi}_p}{M_p} = \sqrt{\frac{4 \lambda_p}{3\vo}}\,\tilde{\tau}_p^{3/4}\,,\qquad 
\frac{\tilde{a}_p}{M_p} = \sqrt{\frac{3 \lambda_p}{4\vo\sqrt{\tilde{\tau}_p}}}\,\tilde{c}_p\,, \nn
\eea
where we did not include the axions $\psi$ and $c_{q_i}$ which are eaten up by two anomalous $U(1)$'s on the D7- and D3-brane stack respectively. Notice that the K\"ahler modulus $T_p = \tau_p + {\rm i} \,c_p$ is given by the following combinations of the canonically normalised fields $\Phi_s = \phi_s + {\rm i}\,a_s$ and $\tilde{\Phi}_p=\tilde{\phi}_p + {\rm i}\,\tilde{a}_p$:
\be
\tau_p = \tilde{\tau}_p- x \tau_s = \left(\frac{3\vo}{4}\right)^{2/3}\left[\frac{1}{\lambda_p^{2/3}}\left(\frac{\tilde{\phi}_p}{M_p}\right)^{4/3} 
- \frac{x}{\lambda_s^{2/3}}\left(\frac{\phi_s}{M_p}\right)^{4/3} \right]\,,
\ee
and:
\be
c_p = \tilde{c}_p- x c_s = \sqrt{\frac{4\vo}{3}}\left(\frac{\tilde{\tau}_p^{1/4}}{\sqrt{\lambda_p}} \frac{\tilde{a}_p}{M_p}  - \frac{x\tau_s^{1/4}}{\sqrt{\lambda_s}} \frac{a_s}{M_p} \right)\,.
\label{cpNorm}
\ee

\subsection{Mass spectrum}
\label{Mass}

The mass matrix around the global minimum and its eigenvalues are derived in detail in App. \ref{AppMass}. Here we just show the leading order volume scaling of the mass of all moduli and charged matter fields for $g_s\sim 0.1$ (in order to trust our approach based on perturbation theory) and $\vo\sim 10^7$. As explained in Sec. \ref{PhenoFeatures}, this choice of the internal volume leads naturally to TeV-scale soft terms for sequestered scenarios with D3-branes at singularities, while it guarantees the absence of any cosmological moduli problem for unsequestered cases with flavour D7-branes. The resulting mass spectrum looks like:

\bea
m_{c_{q_i}} &\sim& m_{\tau_{q_i}} \sim m_\psi \sim m_{|\phi|} \sim M_s \sim g_s^{1/4}\sqrt{\pi}\,\frac{M_p}{\sqrt{\vo}} \sim 10^{15}\,\,\text{GeV}\,, \nn \\
m_{\tau_s} &\sim& m_{c_s} \sim \sqrt{\frac{g_s}{8\pi}}\,\frac{M_p}{\vo}\,\ln\vo \sim 10^{11}\,\,\text{GeV}\,, \nn \\
m_{3/2} &\sim& \sqrt{\frac{g_s}{8\pi}}\,\frac{M_p}{\vo} \sim 10^{10}\,\,\text{GeV}\,, \nn \\
m_{\tilde{\tau}_p} &\sim& \sqrt{\frac{g_s}{8\pi}}\,\frac{M_p}{\vo\sqrt{\ln\vo}} \sim 10^9\,\,\text{GeV}\,, \nn \\
m_{\tau_b} &\sim& \sqrt{\frac{g_s}{8\pi}}\,\frac{M_p}{\vo^{3/2}\sqrt{\ln\vo}} \sim  10^6\,\,\text{GeV}\,, \nn \\
m_{|C|} &\sim& \sqrt{\frac{g_s}{8\pi}} \,\frac{M_p}{\vo^2} \sim 1\,\,\text{TeV}\,, 
\label{masses} \\
m_{c_p} &\sim& \sqrt{\frac{g_s}{8\pi}}\,\frac{M_p}{\vo^{1+p/2}}\,\sqrt{\ln\vo} \sim  10\,\,\text{keV} \qquad\text{for}\quad p = \frac92\,, \nn \\
m_{\theta} &\sim & \frac{\Lambda_{\rm hid}^2}{f_{a_\ALP}} \lesssim 10^{-12}\,\,\text{eV}\,, \nn \\
m_{c_b} &\sim& \sqrt{\frac{g_s}{8\pi}}\frac{M_p}{\vo^{2/3}}\,e^{- \pi\,\vo^{2/3}} \sim 0\,, \nn
\eea
where we focused on the sequestered case with $\alpha_2=2$ illustrated in Sec. \ref{Ffix2} and $\Lambda_{\rm hid}$ represents the scale of strong dynamics in the hidden sector which gives mass to the open string axion $\theta = a_\ALP / f_{a_\ALP}$ whose decay constant is $f_{a_\ALP} = |\hat{C}|\simeq M_p/\vo^2$. As explained in Sec. \ref{PhenoFeatures}, this decay constant leads to a coupling to hidden photons controlled by the scale $M_{\rm hid}\sim 10^6$ GeV that can yield a coupling to ordinary photons via $U(1)$ kinetic mixing given by (\ref{LmixNew}) which can be naturally suppressed by an effective scale of order $M\sim 10^{12}$ GeV. Notice that the DM axion $c_p$ can acquire a mass from poly-instanton effects of order $m_{c_p}\sim 10$ keV if $p=\frac{n}{2}+\kappa =\frac92$, which can be obtained for any $\mc{O}(1)$ value of $n$ by appropriately choosing the flux dependent underlying parameters so that $\kappa \equiv \frac{\tau_p}{\tau_s}=\frac 12 \left(9-n\right)$.

\subsection{DM-ALP coupling}
\label{DM-ALP}

As shown by the mass spectrum in (\ref{masses}) and by the coupling to ordinary photons in (\ref{LmixNew}), the open string axion $\theta$ is a natural candidate for the ALP mode $a_\ALP$ which converts into photons in the magnetic field of galaxy clusters and generates the $3.5$ keV line. However a monochromatic line requires the decay into a pair of ALP particles of a DM particle $a_\DM$ with mass $m_\DM\sim 7$ keV. According to the mass spectrum in (\ref{masses}) $a_\DM$ could be either the local closed string axion $c_p$ or the bulk closed string mode $c_b$ (if $T_b$-dependent non-perturbative effects do not suppress its mass too much). We shall now show that non-perturbative corrections to the matter K\"ahler metric in (\ref{Kmatter}) can induce a coupling of the form $\frac{a_\DM}{\Lambda} \,\partial_\mu a_\ALP \partial^\mu a_\ALP$ due to kinetic mixing between the closed string axion $a_\DM$ and the open string axion $a_\ALP$. We shall also work out the value of the coupling $\Lambda$, finding that it can lie around the Planck/GUT scale only if the DM particle is the local axion $c_p$ ($c_b$ would give a trans-Planckian $\Lambda$). Finally we will explain how in our model a direct DM decay to photons induced by potential couplings of the form $\frac{a_\DM}{4M_\DM}\,F^{\mu\nu}\tilde{F}_{\mu\nu}$ is naturally suppressed by construction. 

In order to compute the DM-ALP coupling, let us focus on contributions to the kinetic Lagrangian of the form:
\be
\mc{L}_{\rm kin}\supset \frac{\partial^2 K}{\partial C \partial \bar{C}}\,\partial_\mu C \partial^\mu \bar{C} 
= \tilde {K}(T_i,\bar{T}_i) \left( \partial_\mu|C|\partial^\mu|C|+ |C|^2\,\partial_\mu\theta\partial^\mu\theta\right) \,.
\label{LkinC}
\ee
If we now expand the closed string axions $c_i$ and the charged open string mode $C=|C|\,e^{{\rm i}\,\theta}$ around the minimum as:
\be
c_i (x) \to \langle c_i\rangle+  c_i(x) \,,\qquad 
|C|(x) \to  \langle |C| \rangle + | C(x)|\,,\qquad \theta(x) \to \langle \theta\rangle + \theta(x) \,,
\ee
the kinetic terms (\ref{LkinC}) become:
\be
\left[ \langle\tilde{K}\rangle - B_i \,e^{-b_i\tau_i}\,\left(\cos(b_i \langle c_i\rangle) \frac{b_i}{2} \hat c_i^2+ \sin(b_i \langle c_i\rangle)\,b_i \hat c_i \right) \right] \left( \partial_\mu|C|\partial^\mu|C|+ |C|^2\,\partial_\mu \theta\partial^\mu \theta\right).
\label{Eq}
\ee
If we now express the open string mode $C$ in terms of the canonically normalised fields $\hat{C}$ and $a_\ALP$ using (\ref{CN}), (\ref{Eq}) contains DM-ALP interaction terms of the form:
\be
\frac{B_i}{2\langle\tilde{K}\rangle} \,e^{-b_i\langle\tau_i\rangle}\,\left(\cos(b_i \langle c_i\rangle) \frac{b_i}{2} \,\hat c_i^2+ \sin(b_i \langle c_i\rangle)\,b_i \,\hat c_i \right) \partial_\mu a_\ALP \partial^\mu a_\ALP\,,
\label{Inter}
\ee
showing that, in order to obtain a three-leg vertex which can induce a two-body DM decay into a pair of ultra-light ALPs, the VEV of $c_i$ has to be such that $b_i\langle c_i\rangle = (2k+1 ) \frac{\pi}{2}$ with $k\in\mathbb{Z}$. Let us therefore focus on this case and consider separately the two options with either $i=b$ or $i=p$: 
\begin{itemize}
\item \textbf{$i=b$ case:} Plugging in (\ref{Inter}) the canonical normalisation for $c_b$ from (\ref{CN}), we find a DM-ALP coupling of the form:
\be
\frac{a_b}{\Lambda}\,\partial_\mu a_\ALP \partial^\mu a_\ALP\qquad\text{with}\qquad \Lambda = \frac{\sqrt{6}\langle\tilde{K}\rangle}{B_b b_b} \,\frac{e^{b_b \langle\tau_b\rangle}}{\langle\tau_b\rangle}\, M_p\sim \displaystyle{
\frac{e^{b_b \vo^{2/3}}}{B_b\, \vo^{4/3}}}\,M_p \gg M_p\,,
\label{Interb}
\ee
which reproduces the value of $\Lambda$ in (\ref{Lambdasize}) for $a_\DM=c_b$. According to the phenomenological constraints discussed in Sec. \ref{PhenoConstr}, $c_b$ cannot play the r\^ole of the DM particle since the scale of its ALP coupling is trans-Planckian. 

\item \textbf{$i=p$ case:} Writing $b_p=\frac{2\pi}{N}$ and using the fact that the minimum for $c_p$ lies at $\langle c_p\rangle = \frac12 + k_1$ with $k_1\in\mathbb{Z}$, the condition $b_p\langle c_p\rangle = (2k_2+1 ) \frac{\pi}{2}$ with $k_2\in\mathbb{Z}$ can be satisfied if $\frac{N}{2} = \frac{(2k_1+1 )}{(2k_2+1 )}$. Hence in the simplest case with $k_1=k_2=0$ we just need $N=2$. Plugging in (\ref{Inter}) the canonical normalisation for $c_p$ from (\ref{cpNorm}), the DM-ALP coupling turns out to be:
\be
\frac{\tilde{a}_p}{\Lambda}\, \partial_\mu a_\ALP \partial^\mu a_\ALP
\qquad\text{with}\qquad \Lambda=  \frac{\sqrt{3\lambda_p}}{\tilde{\tau}_p^{1/4}}\,\frac{\langle\tilde{K}\rangle}{B_p b_p} \,\frac{e^{b_p \langle\tau_p\rangle}}{\sqrt{\vo}}\,M_p \sim \displaystyle{
\frac{M_p}{B_p\, \vo^{7/6-\kappa/N}}} \,,
\label{Interp}
\ee
which reproduces the value of $\Lambda$ in (\ref{Lambdasize}) for $a_\DM= c_p$. This scale of the DM-ALP coupling can easily be around the Planck/GUT scale. For example if $N=2$ and the underlying parameters are chosen such that $\kappa\equiv \tau_p/\tau_s = 2$, $\Lambda \sim M_p /\vo^{1/6}\sim 10^{17}$ GeV for $\vo\sim 10^7$ and $B_p\sim\mc{O}(1)$. Due to the poly-instanton nature of the non-perturbative effects supported by the Wilson divisor $D_p$, the prefactor $B_p$ can however be exponentially small. Comparing $T_p$-dependent poly-instanton corrections to the superpotential in (\ref{Wpoly2}) with $T_p$-dependent non-perturbative corrections to the matter K\"ahler metric in (\ref{Ktilde}), $B_p$ at the minimum could scale as $B_p\sim \mc{O}(\vo^{-1})$. In this case $\Lambda$ can be below the Planck scale only if $\kappa\ll N$. 
\end{itemize}

Let us conclude this section by showing that the branching ratio for direct DM decay into ordinary photons is negligible. Using the fact that the gauge kinetic function for the D7-stack is given by $f_\D7 = T_s + T_p$ (we neglect the flux dependent shift) and the canonical normalisation (\ref{cpNorm}), the closed string axion $c_p = \langle c_p\rangle+\hat{c}_p$ couples to Abelian gauge bosons living on the hidden D7-stack via an interaction term of the form:
\be
\frac{\hat{c}_p}{4\left(\langle \tau_s\rangle + \langle \tau_p\rangle\right)}\,F_{\rm hid}^{\mu\nu}\tilde{F}^{\rm hid}_{\mu\nu} \sim  \frac{\tilde{a}_p}{4 M_s}\,F_{\rm hid}^{\mu\nu}\tilde{F}^{\rm hid}_{\mu\nu}\,.
\ee
One-loop effects generate a kinetic mixing between hidden photons on the D7-stack and ordinary photons on the D3-stack which is controlled by the mixing parameter $\chi\sim 10^{-3}$ given in (\ref{chi}). Thus the DM axion $c_p$ develops an effective coupling to visible sector photons which from (\ref{LmixNew}) looks like:
\be
\frac{\tilde{a}_p}{4\,M_\DM} \,F_{\mu\nu} \tilde{F}^{\mu\nu} \sim \frac{\chi^2\, \tilde{a}_p}{4 M_s}\, F_{\mu\nu} \tilde{F}^{\mu\nu}\qquad\Leftrightarrow\qquad M_\DM \sim \frac{M_s}{\chi^2}\sim 10^5 \,\frac{M_p}{\sqrt{\vo}} \sim 10^{20}\,\,{\rm GeV}\,,
\ee
which is naturally much larger than the scale $\Lambda$ controlling the DM coupling to ALPs.

\section{Conclusions}
\label{Concl}

In this paper we described how to perform a successful global embedding in type IIB string compactifications of the model of \cite{Cicoli:2014bfa} for the recently observed $3.5$ keV line from galaxy clusters. The main feature of this model is the fact that the monochromatic $3.5$ keV line is not generated by the direct decay of a $7$ keV dark matter particle into a pair of photons but it originates from DM decay into ultra-light ALPs which subsequently convert into photons in the cluster magnetic field. Therefore the final signal strength does not depend just on the DM distribution but also on the magnitude of the astrophysical magnetic field and its coherence length which, together with the ALP to photon coupling, determine the probability for ALPs to convert into photons. These additional features make the model of \cite{Cicoli:2014bfa} particularly interesting since it manages to explain not just the observation of a $3.5$ keV line from galaxy clusters but also the morphology of the signal (e.g. the intensity of the line from Perseus seems to be picked at the centre where the magnetic field is in fact more intense) and its non-observation in dwarf spheroidal galaxies (due to the fact that their magnetic field is not very intense and has a relatively small spatial extension). These phenomenological features seem to make this model more promising than standard explanations where DM directly decays into a pair of photons. 

Despite this observational success, the model of \cite{Cicoli:2014bfa} for the $3.5$ keV line did not have a concrete microscopic realisation. In this paper we filled this gap by describing how to construct an explicit type IIB Calabi-Yau compactification which can reproduce all the main phenomenological features of the DM to ALP to photon model. We focused in particular on LVS models since they generically lead to very light axions because some of the moduli are stabilised by perturbative corrections to the effective action. The DM particle is realised as a local closed string axion which develops a tiny mass due to poly-instanton corrections to the superpotential. By an appropriate choice of background and gauge fluxes, the DM mass can be around $7$ keV. The ultra-light ALP is instead given by the phase of an open string mode living on D3-branes at singularities. The ALP decay constant is set by the radial part of this open string mode which is charged under an anomalous $U(1)$. Thus the radial part gets fixed in terms of a moduli-dependent FI-term. In sequestered models with low-energy supersymmetry, the resulting decay constant is naturally in the right ballpark to reproduce a coupling to ordinary photons via $U(1)$ kinetic mixing which is around the intermediate scale, in full agreement with current observations. Notice that future helioscope experiments like IAXO might be able to detect ultra-light ALPs with intermediate scale couplings to photons \cite{Irastorza:2013dav}. Moreover the DM-ALP coupling is generated by kinetic mixing induced by non-perturbative corrections to the K\"ahler potential. For suitable choices of the underlying flux dependent parameters, the scale which controls the associated coupling can be around the GUT/Planck scale, again in good agreement with present observational constraints. 

In this paper we discussed in full depth moduli stabilisation, the mass spectrum and the resulting strength of all relevant couplings but we just described the geometrical and topological conditions on the underlying Calabi-Yau manifold without presenting an explicit example built via toric geometry. This task is beyond the scope of our paper, and so we leave it for future work. Let us however stress that the construction of a concrete Calabi-Yau example with all the desired features for a successful microscopic realisation of our model for the $3.5$ keV line is crucial to have a fully trustworthy scenario. Moreover it would be very interesting to have a more concrete computation of non-perturbative corrections to the 4D $N=1$ K\"ahler potential. 

Another aspect which would deserve further investigation is the cosmological history of our setup from inflation to the present epoch. Here we just point out that the r\^ole of the inflaton could be played by a small blow-up mode like $\tau_s$ \cite{Cicoli:2017shd, Conlon:2005jm}. On the other hand, reheating might be due to the volume mode $\tau_b$ which gets displaced from its minimum during inflation \cite{Cicoli:2016olq} and later on decays giving rise to a reheating temperature of order $T_{\rm rh}\sim 1-10$ GeV \cite{Allahverdi:2013noa}. Such a low reheating temperature would dilute standard thermal WIMP dark matter and reproduce it non-thermally \cite{Allahverdi:2013noa}. Given that in sequestered models with unified gaugino masses the WIMP is generically a Higgsino-like neutralino with an under-abundant non-thermal production in vast regions of the underlying parameter space \cite{Aparicio:2015sda, Aparicio:2016qqb}, an additional DM component in the form of a very light axion like $c_p$ would be needed. Finally one should make sure that tight dark radiation bounds are satisfied since $\tau_b$ could decay both to a pair of ultra-light closed string axions $c_b$ and to a pair of DM axions $c_p$ which could behave as extra neutrino-like species \cite{Cicoli:2012aq, Cicoli:2015bpq}. Notice however that the decay of $\tau_b$ to open string axions $\theta$ living on D3-branes at singularities is negligible since it is highly suppressed by sequestering effects \cite{Cicoli:2012aq}. The DM axions $c_p$ are produced non-thermally at the QCD phase transition via the standard misalignment mechanism. Given that the decay constant of the local closed string axion $c_p$ is of order the string scale which from (\ref{masses}) is rather high, i.e. $M_s \sim 10^{15}$ GeV, axion DM overproduction can be avoided only if the initial misalignment angle is very small. This might be due to a selection effect from the inflationary dynamics \cite{Linde:1991km}. We finally stress that if inflation is driven by a blow-up mode like $\tau_s$, the Hubble scale during inflation is rather low, $H\sim m_{\tau_b} \sim 10^6$ GeV, and so axion isocurvature perturbations would not be in tension with CMB data \cite{Fox:2004kb}. 

\acknowledgments

We would like to thank Joe Conlon, Mark Goodsell and Fernando Quevedo for useful discussions. The work of M.C. is supported by the Programme Rita Levi Montalcini for young researchers of the Italian Ministry of Research.

\appendix

\section{Computational details}
\label{App}

\subsection{Closed string axion decay constants}
\label{AxEFT}

In type IIB string compactifications on Calabi-Yau orientifolds axion-like particles emerge in the low-energy $N=1$ effective field theory from the dimensional reduction of the Ramond-Ramond forms $C_p$ with $p=2,4$. The Kaluza-Klein decomposition under the orientifold projection of these forms is given by \cite{Grimm:2004uq}:
\be
C_2 = c^{i_-}(x)\,\hat{D}_{i_-}\quad \text{and}\quad 
C_4 = c_{i_+}(x) \tilde{D}^{i_+} + Q^{i_+}_2 (x) \wedge \hat{D}_{i_+} + V^{a_+} (x) \wedge \alpha_{a_+} - \tilde{V}_{a_+} (x)\wedge\beta^{a_+}\,, \nn
\ee
where $i_\pm = 1,..., h^{1,1}_\pm$, $a_+=1,...,h^{1,2}_+$, $\tilde{D}^{i_+}$ is a basis of $H^{2,2}_+$ dual to the $(1,1)$-forms $\hat{D}_{i_+}$ and $(\alpha_{a_+}, \beta^{a_+})$ is a real, symplectic basis of $H^3_+ = H^{1,2}_+ \oplus H^{2,1}_+$. 

As explained in Sec. \ref{Dfix}, in our model the orientifold-odd axions $c_{i_-}$, if present, are eaten up by anomalous $U(1)$'s in the process of anomaly cancellation. We shall therefore focus on the case with $h^{1,1}_-=0$ where the K\"ahler moduli take the simple expression $T_i = \tau_i + {\rm i} \,c_i$ with $i=1,...,h^{1,1}_+ = h^{1,1}$. 

The coupling of orientifold-even closed string axions to $F\wedge F$ can be derived from the Kaluza-Klein reduction of the Chern-Simons term of the D-brane action. Moreover, the periods of the canonically unnormalised axions $c_i$ are integer multiples of $M_p$ and their kinetic terms read \citep{Cicoli:2012sz}:
\be
\mc{L}_{\rm kin} = K_{ij} \partial_\mu c_i \partial^\mu c_j = \frac18 \,\eta_i\, \partial_\mu c'_i \partial^\mu c'_i\,,
\ee
where the $c'_i$'s are the axions which diagonalise the K\"ahler metric $K_{ij}$ and $\eta_i$ are its eigenvalues. A proper canonical normalisation of the kinetic terms can then be easily obtained by defining: 
\be
\frac18 \,\eta_i \partial_\mu c'_i \partial^\mu c'_i \equiv \frac12 \,\partial_\mu a_i \partial^\mu a_i \qquad \text{with}\qquad 
a_i = \frac12 \,\sqrt{\eta_i} \,c'_i\,,
\ee
which shows that the canonically normalised axions $a_i$ acquire periods of the form:
\be
\frac{2}{\sqrt{\eta_i}} \,a_i = \frac{2}{\sqrt{\eta_i}} \,a_i + M_p\qquad \Rightarrow\qquad 
a_i = a_i + \frac{\sqrt{\eta_i}}{2}\, M_p\,. 
\ee
We can then set the conventional axionic period as:
\be
a_i = a_i + 2\pi f_{a_i} \qquad \text{with}\qquad f_{a_i} =\frac{\sqrt{\eta_i}\,M_p}{4\pi}\,,
\ee 
where $f_{a_i}$ is the standard axion decay constant. Closed string axions which propagate in the bulk of the extra dimensions have a decay constant of order the Kaluza-Klein scale $M_\KK \sim M_p/\vo^{2/3}$, whereas the decay constant of closed string axions whose corresponding saxion parameterises the volume of localised blow-up modes is controlled by the string scale $M_s \sim M_p /\sqrt{\vo}$:
\be
\label{BulkLocAx}
 f_{a_i}\simeq\left\lbrace 
\begin{array}{cc}
M_p/\tau_i \sim M_\KK & \qquad\text{bulk axion} \\ 
M_p/\sqrt{\vo}\sim M_s & \qquad\text{local axion}
\end{array}\right.
\ee
Notice however that the axion coupling to the Abelian gauge bosons living on the D-brane wrapping the four-cycle whose volume is controlled by the associated saxion $\tau_i$, is given by:
\be
\frac{g_i^2}{32\pi^2}\,\frac{a_i}{f_{a_i}}\,F^{(i)}_{\mu\nu}\tilde{F}_{(i)}^{\mu\nu} = \frac{1}{32\pi^2}\,\frac{a_i}{\tau_i f_{a_i}}\,F^{(i)}_{\mu\nu}\tilde{F}_{(i)}^{\mu\nu} \,,
\label{AxCoupl}
\ee
since the gauge coupling is set by the saxion as $g_i^2=\tau_i$. Hence combining (\ref{BulkLocAx}) with (\ref{AxCoupl}) we realise that that the coupling of bulk closed string axions to gauge bosons is controlled by $M \sim \tau_i f_{a_i} \sim M_p$, in agreement with the fact that moduli couple to ordinary matter with gravitational strength. On the other hand the coupling of local closed string axions to gauge bosons is set by the string scale $M_s$ which in LVS models with exponentially large volume can be considerably smaller than the Planck scale.

\subsection{Canonical normalisation}
\label{AppCanNorm}

The kinetic terms for all K\"ahler moduli and the charged open string modes $\phi$ and $C$ can be derived from the total K\"ahler potential $K = K_{\rm mod} + K_{\rm matter}$, where $K_{\rm mod}$ is given by the three contributions in (\ref{K}) and $K_{\rm matter}$ is shown in (\ref{Kmatter}) and (\ref{Ktilde}), as follows:
\be
\mc{L}_{\rm kin} = \frac{\partial^2 K}{\partial \chi_i \partial \bar{\chi}_{\bar{j}}} \,\partial_\mu \chi_i \partial^\mu \bar{\chi}_{\bar{j}} \,,
\ee
where $\chi_i$ denotes an arbitrary scalar field of our model. As can be seen from (\ref{Kmatter}), the D7 open string mode $\phi$ mixes only with the dilaton $S$, and so can be easily written in terms of the corresponding canonically normalised field $\hat\phi$ as:
\be
\frac{\hat\phi}{M_p} = \sqrt{\frac{2}{{\rm Re}(S)}}\, \phi\,.
\ee
From the first term in (\ref{K}) we also realise that cross-terms between the blow-up mode $\tau_{q_i}$ and any of the other K\"ahler moduli are highly suppressed when evaluated at the minimum for $\tau_{q_i}\simeq 0$ (more precisely, as discussed in Sec. \ref{Ffix2}, depending on the level of sequestering of soft masses, we can have either $\tau_{q_i}\sim \vo^{-1}\ll 1$ or $\tau_{q_i}\sim \vo^{-3}\ll 1$). Hence it is straightforward to write also $\tau_{q_i}$ in terms of the corresponding canonically normalised field $\phi_{q_i}$ as:
\be
\frac{\phi_{q_i}}{M_p} = \frac{\tau_{q_i}}{\sqrt{\vo}} \qquad \text{for} \quad i=1,2\,.
\ee
The remaining fields $T_b$, $T_s$, $T_p$ and $C$ mix with each other, leading to a non-trivial K\"ahler metric whose components take the following leading order expressions for $\vo \simeq \lambda_b \tau_b^{3/2} \gg 1$:
\bea
K_{T_i \bar{T}_{\bar{j}}} &\simeq& \frac{3}{8\vo}\left(
\begin{array}{ccc}
\frac{2 \lambda_b}{\sqrt{\tau_b}} & -\frac{3}{\tau_b}\left(\lambda_s\sqrt{\tau_s}+x\lambda_p\sqrt{\tilde{\tau}_p}\right) & 
-\frac{3}{\tau_b}\,\lambda_p\sqrt{\tilde{\tau}_p} \\ 
-\frac{3}{\tau_b}\left(\lambda_s\sqrt{\tau_s}+x\lambda_p\sqrt{\tilde{\tau}_p}\right) & \frac{\lambda_s}{\sqrt{\tau_s}} + \frac{x^2 \lambda_p}{\sqrt{\tilde{\tau}_p}} & 
\frac{x \lambda_p}{\sqrt{\tilde{\tau}_p}} \\ 
-\frac{3}{\tau_b}\,\lambda_p\sqrt{\tilde{\tau}_p} & \frac{x \lambda_p}{\sqrt{\tilde{\tau}_p}} & \frac{\lambda_p}{\sqrt{\tilde{\tau}_p}} 
\end{array}
\right) \nn \\
K_{T_b\bar{C}}&\simeq&-\frac{\tilde{K}}{2\,\tau_b}\,C\,, \qquad
K_{T_s \bar{C}}\simeq \frac{\tilde{K}}{2\vo}\left(\lambda_s\sqrt{\tau_s} + x \lambda_ p \sqrt{\tilde{\tau}_p}\right)C\,, \nn \\
K_{T_p \bar{C}} &\simeq& \frac{\tilde{K}}{2\vo}\,\lambda_p\sqrt{\tilde{\tau}_p}\,C\,,\qquad
K_{C\bar{C}} =  \tilde{K}\,.
\nn
\eea
In the large volume limit, different contributions to the kinetic Lagrangian can be organised in an expansion in $1/\vo\ll 1$ as follows:
\be
\mc{L}_{\rm kin} = \mc{L}_{\rm kin}^{\mc{O}(1)}+\mc{L}_{\rm kin}^{\mc{O}(\vo^{-1})}+\mc{L}_{\rm kin}^{\mc{O}(\vo^{-4/3})}\,, \nn
\ee
where, trading $T_p$ for $\tilde{T}_p = T_p+x T_s$, we have:
\bea
\mc{L}_{\rm kin}^{\mc{O}(1)} &=& \frac{3}{4\,\tau_b^2}\,\partial_\mu \tau_b \partial^\mu\tau_b \,, \nn \\
\mc{L}_{\rm kin}^{\mc{O}(\vo^{-1})} &=& \frac{3}{8\vo}\left[\frac{\lambda_s}{\sqrt{\tau_s}}\left(\partial_\mu \tau_s \partial^\mu\tau_s+\partial_\mu c_s\partial^\mu c_s\right)+\frac{\lambda_p}{\sqrt{\tilde{\tau}_p}}\left(\partial_\mu \tilde{\tau}_p \partial^\mu\tilde{\tau}_p +\partial_\mu \tilde{c}_p\partial^\mu \tilde{c}_p\right)\right] \nn \\
&-& \frac{9}{4\vo}\frac{\partial_\mu \tau_b}{\tau_b}\left( \lambda_s\sqrt{\tau_s}\,\partial^\mu\tau_s +\lambda_p\sqrt{\tilde{\tau}_p}\,\partial^\mu\tilde{\tau}_p \right)\,, \nn \\
\mc{L}_{\rm kin}^{\mc{O}(\vo^{-4/3})} &=& \frac{3}{4\,\tau_b^2}\,\partial_\mu c_b\partial^\mu c_b\,. \nn
\eea
At leading order the kinetic terms become canonical if $\tau_b$ is replaced by $\phi_b$ defined as: 
\be
\frac{\phi_b}{M_p} = \sqrt{\frac32}\,\ln\tau_b\,,
\label{phib}
\ee
whereas $\mc{L}_{\rm kin}^{\mc{O}(\vo^{-1})}$ becomes diagonal if the small modulus $T_s$ and the Wilson modulus $\tilde{T}_p$ are substituted by:
\bea
\frac{\phi_s}{M_p} &=& \sqrt{\frac{4 \lambda_s}{3\vo}}\,\tau_s^{3/4}\,, \qquad
\frac{a_s}{M_p} = \sqrt{\frac{3 \lambda_s}{4\vo\sqrt{\tau_s}}}\,c_s\,,   \nn \\
\frac{\tilde{\phi}_p}{M_p} &=& \sqrt{\frac{4 \lambda_p}{3\vo}}\,\tilde{\tau}_p^{3/4}\,,\qquad 
\frac{\tilde{a}_p}{M_p} = \sqrt{\frac{3 \lambda_p}{4\vo\sqrt{\tilde{\tau}_p}}}\,\tilde{c}_p\,,
\label{CNorm}
\eea
and the canonical normalisation (\ref{phib}) for $\tau_b$ gets modified by the inclusion of a subleading mixing with $\tau_s$ and $\tilde{\tau}_p$ of the form:
\be
\frac{\phi_b}{M_p} = \sqrt{\frac32}\,\ln\tau_b-\sqrt{\frac23}\,\frac{1}{\vo}\left(\lambda_s\tau_s^{3/2}+\lambda_p\tilde{\tau}_p^{3/2}\right)\,.
\label{phibmix}
\ee
Finally the kinetic term in $\mc{L}_{\rm kin}^{\mc{O}(\vo^{-4/3})}$ are canonically normalised if the bulk axion $c_b$ gets redefined as:
\be
\frac{a_b}{M_p} = \sqrt{\frac32} \frac{c_b}{\tau_b}\,.
\ee
The $U(1)$-charged open string mode $C$ appears in the kinetic Lagrangian only at $\mc{O}(|C|^2\vo^{-2/3})$ which according to (\ref{fa1}) and (\ref{fa2}) can scale as either $\vo^{-8/3}$ or $\vo^{-14/3}$. This part of the kinetic Lagrangian looks like:
\be
\mc{L}_{\rm kin}\supset \tilde{K}\,|C|^2\left(\frac{\partial_\mu |C|}{|C|}\frac{\partial^\mu |C|}{|C|} + \partial_\mu\theta\partial^\mu\theta 
- \frac{\partial_\mu \tau_b}{\tau_b}\frac{\partial^\mu |C|}{|C|}\right)\,,
\ee
and becomes diagonal by redefining: 
\be
\frac{|\hat{C}|}{M_p} = \sqrt{2\tilde{K}} |C| \qquad \text{and}\qquad a_\ALP = |\hat{C}| \theta  = f_{a_\ALP} \theta\,. \nn 
\ee

\subsection{Mass matrix}
\label{AppMass}

As described in Sec. \ref{Dfix}, the moduli stabilised at tree-level are $\tau_{q_i}$ and $|\phi|$ while the corresponding axions are eaten up by two anomalous $U(1)$'s. Given that they fixed at $\mc{O}(1/\vo^2)$, all these modes develop a mass of order the string scale: 
\be
m_{\tau_{q_i}}\sim m_{c_{q_i}}\sim m_{|\phi|}\sim m_{\psi} \sim M_s = g_s^{1/4} \sqrt{\pi}\,\frac{M_p}{\sqrt{\vo}}\,.
\ee
On the other hand, $\tau_b$, $\tau_s$, $\tilde{\tau}_p$ and the closed string axion $c_s$ are stabilised at $\mc{O}(1/\vo^3)$. The masses of the corresponding canonically normalised fields derived in App. \ref{AppCanNorm} are given by the eigenvalues of the mass matrix evaluated at the minimum of the  $\mc{O}(1/\vo^3)$ scalar potential. The leading order contributions of all the elements of this $4\times 4$ matrix read:
\bea
\frac{\partial^2 V}{\partial \phi_b\partial \phi_b} &=& \left(\frac{g_s}{8\pi}\right) \frac{9\,\lambda_s\,\tau_s^{3/2}}{2}
\frac{W_0^2}{\vo^3}\,,  \nn \\
\frac{\partial^2 V}{\partial \phi_b\partial \phi_s} &=& \left(\frac{g_s}{8\pi}\right) \frac{3\sqrt{2\lambda_s}\,\tau_s^{3/4}}{\sqrt{\vo}} \left(\frac{W_0}{\vo}\right)^2 \left(2\pi\tau_s\right), \nn \\
\frac{\partial^2 V}{\partial \phi_s\partial \phi_s} &=& \frac{\partial^2 V}{\partial a_s \partial a_s} = 4 \left(\frac{g_s}{8\pi}\right) \left(\frac{W_0}{\vo}\right)^2 \left(2\pi\tau_s\right)^2, \nn \\
\frac{\partial^2 V}{\partial \tilde{\phi}_p\partial \tilde{\phi}_p} &=& \left(\frac{g_s}{8\pi}\right) \frac{1}{4 z_p \tilde{\tau}_p}\left(\frac{W_0}{\vo}\right)^2 , \nn  \\
\frac{\partial^2 V}{\partial \phi_b \partial \tilde{\phi}_p} &=& \frac{\partial^2 V}{\partial \phi_b\partial a_s} = \frac{\partial^2 V}{\partial \phi_s\partial \tilde{\phi}_p} = \frac{\partial^2 V}{\partial \phi_s \partial a_s} = \frac{\partial^2 V}{\partial \tilde{\phi}_p\partial a_s} = 0\,, \nn
\eea
The eigenvalues of this mass matrix turn out to be:
\bea
m_{\phi_s}^2 &=& m_{a_s}^2  = 4 \left(\frac{g_s}{8\pi}\right) \left(\frac{W_0}{\vo}\right)^2 \left(2\pi\tau_s\right)^2 \simeq m_{3/2}^2 \left(\ln\vo\right)^2\,, \nn \\
m^2_{\tilde{\phi}_p} &=& \left(\frac{g_s}{8\pi}\right) \frac{\pi}{2 z_p}\left(\frac{W_0}{\vo}\right)^2 \frac{1}{2\pi\tilde{\tau}_p} \simeq \frac{m_{3/2}^2}{\ln\vo}
\qquad\text{and}\qquad m_{\phi_b}^2 =0\,,
\eea
where the gravitino mass is given by:
\be
m_{3/2}^2 = e^K\,|W|^2\simeq \left(\frac{g_s}{8\pi}\right) \left(\frac{W_0}{\vo}\right)^2 \,.
\ee
The mass of the canonically normalised large modulus $\phi_b$ becomes non-zero once we include subleading $1/(2\pi\tau_s)\sim 1/\ln\vo\ll 1$ corrections to the elements of the mass matrix, and scales as (with $c$ an $\mc{O}(1)$ numerical coefficient):
\be
m_{\phi_b}^2 =  c\,\lambda_s \tau_s^{3/2} \left(\frac{g_s}{8\pi}\right) \frac{W_0^2}{\vo^3} \frac{1}{2\pi\tau_s}\simeq \frac{m_{3/2}^2}{\vo\ln\vo}\,.
\ee
As explained in Sec. \ref{Ffix2}, the charged matter field $|C|$ is fixed by soft supersymmetry breaking contributions to the scalar potential and can acquire a mass of order $m_{3/2}/\sqrt{\vo}$ or $m_{3/2}/\vo$ depending on the level of sequestering. The corresponding phase $\theta = a_\ALP/f_{a_\ALP}$ behaves as an open string ALP which develops a mass of order: 
\be
m_{a_\ALP} \sim \frac{\Lambda_{\rm hid}^2}{f_{a_\ALP}} \sim \frac{\Lambda_{\rm hid}^2}{|\hat{C}|}\,,
\ee
where $\Lambda_{\rm hid}$ is the scale of strong dynamics effects in the hidden sector. In order to obtain a phenomenologically viable value $m_{a_\ALP}\lesssim 10^{-12}$ eV, we need to have $\Lambda_{\rm hid}\lesssim 10^4$ eV if $f_{a_\ALP}\sim m_{3/2}\sim 10^{10}$ GeV or $\Lambda_{\rm hid}\lesssim 1$ eV if $f_{a_\ALP}\sim m_{3/2}/\vo \sim 1$ TeV. 

The DM axion $c_p$ is stabilised by tiny poly-instanton corrections at $\mc{O}(1/\vo^{3+p})$. Using the fact that $K^{-1}_{T_p \bar{T}_p}\sim \vo\sqrt{\tilde{\tau}_p}$ and the expression (\ref{Vpoly}) for the scalar potential for $c_p$, its mass can be easily estimated as:
\be
m^2_{\tilde{a}_p} \sim K^{-1}_{T_p \bar{T}_p} \frac{\partial^2 V_F^{\rm poly}(c_p)}{\partial c_p^2}  \sim 
\left(\frac{g_s}{8\pi}\right)\, \frac{W_0^2}{\vo^{2+p}}\,2\pi\tilde{\tau}_p\sim \frac{m_{3/2}^2}{\vo^p}\,\ln\vo\,.
\label{csmass}
\ee
If the volume is of order $\vo\sim 10^7$, this mass can be around $10$ keV if $p=9/2$. As explained in Sec. \ref{Mass} this value of $p$ can be accommodated by an appropriate choice of underlying flux parameters. Finally the axion $c_b$ of the large modulus $T_b = \tau_b + {\rm i}\,c_b$ can receive a potential only from highly suppressed non-perturbative contributions to the superpotential of the form $W_{\rm np}\supset A_b \,e^{-2\pi T_b}$ which can be shown to lead to a mass for the axion $c_b$ that scales as:
\be
m_{a_b}^2 \sim \left(\frac{g_s}{8\pi}\right)\,\frac{M_p^2}{\vo^{4/3}}\,e^{-\frac{2\pi}{\lambda_b^{2/3}}\, \vo^{2/3}} \sim 0\,.
\ee

\bibliographystyle{JHEP}

\end{document}